\documentclass[a4paper,11pt]{article}
\pdfoutput=1 % if you are submitting a pdflatex (i.e. if you have
\usepackage{jcappub} % for details on the use of the package, please
\usepackage[T1]{fontenc} % if needed
\usepackage[utf8]{inputenc}
\usepackage{float}
\usepackage{caption}
\usepackage{subcaption}
\usepackage{placeins}

\DeclareUnicodeCharacter{2212}{-}

\title{\boldmath Oscillon Formation in Palatini Modified Gravity Theories}

\author[a]{Shreyas Upadhye}
\author[a]{and Sukanta Panda}
\affiliation[a]{Department of Physics, Indian Institute of Science Education and Research, Bhopal, India, 462066}

\emailAdd{shreyas.upadhye26@gmail.com}
\emailAdd{sukanta@iiserb.ac.in}

\abstract{We investigate the formation of spatially localized, time-oscillatory, non-topological solitonic, and quasi-stable energy configurations known as oscillons. These structures are formed at the end of the inflationary epoch during the preheating phase and decay over long periods of time. Oscillons have been previously studied in the literature within the regime of General Relativity using the metric formalism. In this paper, we investigate the formation of these energy lumps by modifying the gravity sector of the Einstein-Hilbert action. We consider a non-minimal coupling of the scalar field to the Ricci scalar, $R$, and work within an alternative formulation of General Relativity known as the \textit{Palatini formalism}. The scalar potential we consider takes a polynomial form. Using \textit{Cosmo}$\mathcal{L}$\textit{attice}, we demonstrate numerically that the equation governing the dynamics of the inflaton scalar field yields the oscillatory and decaying solutions expected for oscillons, with the power spectrum governing the growth of $k$-mode perturbations. The equation of state reveals an extended period of oscillon domination in the early universe. Additionally, the primordial gravitational wave spectrum generated by the asymmetric distribution of these energy configurations has been studied. We observe that these dynamics generate gravitational waves in the ultra-high-frequency regime, which lie within the sensitivity range of planned future detectors and experiments.}

\begin{document}
\maketitle
% REMOVED \flushbottom FROM HERE

\section{Introduction}
\label{sec:intro}
The theory of Cosmic Inflation \cite{PhysRevD.23.347,LINDE1982389,LINDE1983177} is the most widely accepted, as it provided the initial conditions required to tackle the conceptual problems of the hot Big Bang while generating the primordial density fluctuations seeding the large-scale structure formation we see today. This has been backed by the constant improvement in the precision of measurements of the Cosmic Microwave Background (CMB) data \cite{Planck:2018jri,Ade2016,Ade_2021}. We achieve the Standard Model of Cosmology by integrating this phase of the early universe, where we have accelerated expansion, with late-time expansion driven by dark energy, which is described by the $\Lambda CDM$ model. Beyond driving exponential expansion, a viable inflationary model must incorporate a consistent `graceful exit' mechanism. This mechanism facilitates the non-perturbative decay of the homogeneous inflaton condensate, effectively transferring its energy density into Standard Model (SM) degrees of freedom. This preheating phase is essential for the thermalization of the universe, marking the formal onset of the hot Big Bang cosmology \cite{RevModPhys.78.537,Allahverdi_2010}.
The so-called preheating phase \cite{Kaiser_1997,Kofman_1997,Greene_1997}, also known as the initial stage of reheating, is typically quite non-linear and non-perturbative, demanding the employment of numerical techniques like classical lattice simulations \cite{Figueroa_2021}.

It is widely recognized that the rotation curves of spiral galaxies have a non-Keplerian circular velocity profile that cannot be explained by a Newtonian gravitational potential resulting from baryonic matter \cite{1996MNRAS.281...27P}. To resolve this problem, there have been several proposed explanations, with dark matter being the most popular one, and it is a key ingredient of the $\Lambda$CDM model. Nevertheless, it is met with significant challenges \cite{Famaey_2012,2012PASA...29..395K,Bullock_2017,Di_Valentino_2021}, which prompted the need to introduce a phenomenological modification to Newtonian dynamics and higher-order gravitational theories, thus opening the domain of modified gravity, which also provided an alternative to dark energy \cite{Carroll_2005}. 
This resulted in a number of gravity theories being formulated, for example, DGP (Dvali-Gabadadze-Porrati) gravity \cite{Dvali_2000}, brane-world gravity \cite{Maartens_2010}, TeVeS (Tensor-Vector-Scalar), and Einstein-Aether theory \cite{Jacobson_2001}. The most well-known is the scalar-tensor theory \cite{PhysRevD.1.3209,Bergmann1968}.

One such widely accepted class of theories, which can be considered as the simplest modification, is the $f(R)$ theory of gravity \cite{Sotiriou_2010, De_Felice_2010,Nojiri_2011,Nojiri_2017}, where the Einstein-Hilbert action is written in terms of an arbitrary function of the Ricci scalar,
\begin{equation}
     S_{EH}=\frac{1}{2\kappa}\int d^4x\sqrt{-g} \hspace{0.1cm}f(R)
     \label{1.2}
\end{equation}
For the standard General Relativity case, we have $f(R)=R$ and the action takes the form of the usual Einstein-Hilbert action. The $f(R)$ set of theories is easy to cope with and deal with, aside from its inspiration from high-energy physics. Additionally, these theories dodge the deadly Ostrogradski instability \cite{Woodard2007}, making them unique in comparison to other higher-order theories. When dealing with such a class of theories, the choice of the variational principle becomes crucial. Although various formalisms can be employed to derive field equations from an action—depending on the underlying theoretical framework—the most prevalent is the metric formalism. In this approach, the connection is assumed to be the Levi-Civita connection, which is uniquely determined by the metric tensor. Consequently, the field equations are obtained by varying the Einstein-Hilbert action solely with respect to the metric $g_{\mu\nu}$. In this work, we adopt the \textit{Palatini formalism}, a framework particularly advantageous when investigating modified gravity actions. Within this approach, the metric $g_{\mu\nu}$ and the connection $\Gamma^{\lambda}_{\mu\nu}$ are treated as independent dynamical variables, with the field equations derived by varying the action with respect to both. A defining feature of this formalism is that the matter action is assumed to be independent of the connection, which is essential for recovering the standard Einstein field equations in the appropriate limit. Indeed, for the Hilbert action where $f(R) = R$, the metric and Palatini formulations yield identical physical results. When we study the $f(R)$ theories of gravity in the Palatini formalism, there are no solutions possible that give rise to inflation due to the absence of additional propagating degrees of freedom, unlike in the metric formalism where they are present. Hence, a scalar field needs to be added to the action to study inflation in the Palatini formalism \cite{Panda_2023}, and we enter the regime of scalar-tensor theories. Inflation in $f(R)$ theories in the Palatini formalism has been well studied in \cite{Enckell_2019,2018JCAP...11..028A,BAUER2008222,Annala:2020cqj,sym12121958}. The current cosmic acceleration owing to dark energy was studied in the Palatini formalism in \cite{Allemandi_2005}. The preference for the Palatini formalism over the metric formalism arises from the straightforward solutions it offers and a more general structure.

Generic nonlinear scalar field theories can support spatially localized field configurations that undergo sustained oscillations in time \cite{dymnikova2000quasilumpsorderphasetransitions}. Such solutions typically arise in models where the scalar potential becomes shallower than a purely quadratic form ($\frac{m^2\phi^2}{2}$) away from its minimum. In this case, as the inflaton oscillates about the minimum of the potential, it experiences attractive self-interactions. These attractive interactions enable the formation of localized, long-lived, and spatially dense field configurations known as \textbf{oscillons} \cite{Gleiser_1994,Kusenko_1997,Kasuya_2003,Farhi_2008,Sang_2021}.
Oscillons may have major implications for the post-inflationary universe. Long after the homogeneous inflaton has broken up, they could result in an accelerated inflaton decay rate or explosive particle generation in specific areas. On considerably smaller scales than those responsible for the current large-scale structure, enhanced density perturbations caused by oscillons serve as seeds for the creation of structures through mergers. Primordial black holes may be produced from large-sized oscillons \cite{Cotner_2018,Cotner_2019,Widdicombe_2020,Nazari_2021,khlopov1999orderphasetransitionssource}. It will be interesting to study the gravitational wave production due to the formation, collapse, and merger of these lumps \cite{Zhou:2013tsa,Easther_2006,Liu_2018,Antusch_2017}.

The features of oscillons suggest that they can affect the standard picture of scalar ultra-light dark matter models \cite{Oll__2020,Kawasaki_2020,Ferreira_2021}. Oscillons have been well studied in the literature in the regime of General Relativity \cite{amin2010inflatonfragmentationemergencepseudostable,Amin_2010,Amin_2010f,Amin:2011hj}. In this work, we try to probe numerically the solutions that give rise to oscillons in the preheating era of the universe, starting just after inflation.

We examine a modified gravity model in the Palatini formalism where we have a non-minimal coupling of the scalar field to the curvature inspired by the Starobinsky model of inflation \cite{STAROBINSKY198099}, $\mathcal{L} \propto (1+\zeta \frac{\Phi^2}{M^2_{pl}})(R+\alpha R^2)$. The potential we take into consideration is an attractive, non-linear, 6th-order polynomial function of the inflaton field $\Phi$. We perform classical lattice simulations using the publicly available \textit{Cosmo}$\mathcal{L}$\textit{attice} package \cite{Figueroa_2023} to extract the energy density, power spectrum, cosmological equation of state parameter, and the spectra of gravitational waves generated by oscillons.

The paper is organized as follows: In Section \ref{sec : The model}, we build up the mathematical framework of the model under consideration with an overview of the action, pressure, energy density, and subsequently the energy-momentum tensor. Section \ref{sec: linear stability} involves the study of Floquet theory, which governs the growth of perturbations. In Section \ref{sec:numerical}, we adopt a full numerical approach and explore the power spectrum, mean field values, state of the universe during oscillon formation, and the gravitational wave spectrum generated from oscillons. We conclude our work by providing a summary in Section \ref{Sec: summary}. The detailed calculation of the Floquet exponents is given in the Appendix.

Throughout the paper, we mostly work in natural units considering $c = \hbar = 1$ and $M_{pl}=1$ (reduced Planck mass) for analytic computation. But while performing lattice simulations, we take $M_{pl}=2.44\times10^{18} \text{ GeV}$. The signature of our flat FLRW metric is positive $(-,+,+,+)$.

\section{The Model}
\label{sec : The model}
\subsection{Action}
We formulate a Palatini inflation model in $f(R,\Phi)$ theory, where $\Phi$ is the inflaton scalar field and $R$ is the curvature scalar \cite{Das_2021, Kuralkar:2025hoz}. As of now, we do not define the explicit form for $f(R,\Phi)$ and keep it general, but it is a non-linear function of $R$. We start with a general action in the Jordan frame of the form:
\begin{equation}
    S=\int  \sqrt{-g}[\frac{1}{2}f(R,\Phi)-\frac{1}{2}g^{\mu\nu}\partial_\mu\Phi \partial_\nu \Phi-V(\Phi)]d^4x+S_{m}(g_{\mu\nu},\psi).
    \label{2.1}
\end{equation}
where $g_{\mu\nu}$ is the metric and $g$ is its determinant, $R$ is the Ricci Scalar defined as $R=g^{\alpha\beta}R_{\alpha\beta}(\Gamma)$ in Palatini formalism. The matter action $S_m$ consists of the other matter components of the universe. 

The non linearity of $f(R,\Phi)$ makes the gravitational equations of motion complicated to simplify it we introduce an auxiliary field $\phi$ to linearize the action in R and this brings us to the Scalar-Tensor theory which modifies the action as follows:
\begin{equation}
    S = \int d^4x\sqrt{-g} \left[ \frac{1}{2} f(\phi, \Phi) + \frac{1}{2} f'(\phi, \Phi)(R - \phi) - \frac{1}{2} g^{\mu\nu} \partial_{\mu}\Phi \partial_{\nu}\Phi - V(\Phi) \right]
    \label{2.2}
\end{equation}
where $f'(\phi,\Phi)=\frac{\partial f(\phi,\Phi)}{\partial \phi}$. 

As the auxiliary field, $\phi$, wont generate any kinetic terms so it would also not produce any equations of motion, instead it would give a constraint equation if we vary the action with respect to $\phi$ :
\begin{equation}
    \delta \mathbf{S} = 0 \implies \phi = R \quad \text{for} \hspace{0.5cm} \frac{\partial^2f(\phi,\Phi)}{\partial \phi^2} \ne 0 
    \label{2.3}
\end{equation}
Using this in \eqref{2.2} we recover our original action \eqref{2.1}. Rearranging \eqref{2.2} gives : 
\begin{equation}
    S = \int d^4x\sqrt{-g} \left[ \frac{1}{2} f'(\phi, \Phi)R - W(\phi, \Phi) - \frac{1}{2}g^{\mu\nu}\partial_{\mu}\Phi\partial_{\nu}\Phi - V(\Phi) \right]
    \label{2.4}
\end{equation}
where $ W(\phi, \Phi)=\frac{1}{2}\phi f'(\phi,\Phi)-\frac{1}{2}f(\phi,\Phi)$. We emphasize on the epoch where the coupling between the inflaton field and matter (radiation) is weak  \cite{Dimopoulos_2022}. So we can effectively ignore the matter action in \eqref{2.1}.
In this case, a single scalar field $\Phi$, that is non-minimally coupled to gravity is described by the action \eqref{2.4}. The Weyl (conformal) transformation \cite{10.1063/1.1664478} is used to produce a minimally coupled scalar field in the Einstein frame. The transformation can be expressed as follows and is dependent on both $\Phi$ and $\phi$ 
\begin{equation}
     g_{\mu\nu} \xrightarrow{}f'(\phi,\Phi)g_{\mu\nu}
     \label{2.5}
\end{equation}
Applying this transformation the action obtained in the Einstein frame can be written as : 
\begin{equation}
    S = \int d^4x\sqrt{-g} \left[ \frac{1}{2}R - \frac{1}{2f'(\phi, \Phi)}\partial_{\mu}\Phi\partial^{\mu}\Phi - \frac{W(\phi, \Phi) + V(\Phi)}{f'(\phi, \Phi)^2} \right]
    \label{2.6}
\end{equation}
Let us define a new potential :
\begin{equation}
    \hat{V}(\phi,\Phi)= \frac{W(\phi,\Phi)+V(\Phi)}{f'(\phi,\Phi)^2}
    \label{2.7}
\end{equation}
Here we consider our coupling function to be $f(R,\Phi)=G(\Phi)(R+\alpha R^2)$. Accordingly the potential gets modified as:
\begin{equation}
    \hat{V}=\frac{1}{f'(\phi,\Phi)^2}\left[\frac{1}{8\alpha G(\Phi)}[f'(\phi,\Phi)-G(\Phi)]^2+V(\Phi)\right]
    \label{2.8}
\end{equation}
Now varying the action equation~\eqref{2.6} with respect to $\phi$ and equating to $0$ we get the new constraint equation to be:
\begin{equation}
    f'(\phi,\Phi)=\frac{8\alpha V(\Phi)+G(\Phi)}{1-2\alpha \partial_\mu\Phi\partial^\mu\Phi}
    \label{2.9}
\end{equation} 
Substituting equation~\eqref{2.8} and ~\eqref{2.9} into the action equation~\eqref{2.6}, we get the new action to be
\begin{equation}
    S=\int d^4x \sqrt{-g}\left[\frac{R}{2}-\frac{X}{(8\alpha V+G)}+X^2\frac{2\alpha}{(8\alpha V+G)}-\frac{V}{G(8\alpha V+G)}\right]
    \label{2.10}
\end{equation}
where $X=-\frac{1}{2}\partial_\mu\Phi\partial^\mu\Phi$. The potential is taken to be $V(\Phi)=\frac{m^2\Phi^2}{2}-\frac{\lambda\Phi^4}{4}+\frac{\sigma\Phi^6}{6m^2}$ (the reason for this particular choice is described in section \ref{sec:numerical} and the coupling function is $G(\Phi)=1+\zeta\Phi^2$.

\vspace{0.3cm} 
The values of the parameters are fixed as: $m=5 \times10^{-6} M_{pl}$, $\lambda=1 \times10^{-7}$, $\sigma=1 \times10^{-14}$, $\alpha=0.005$, and $\zeta= 50000$. The mass parameter $m$ is consistent with the CMB normalization of the scalar power spectrum $\mathcal{P}_s \sim 2 \times 10^{-9}$ for inflation near the plateau of $U(\chi)$. The quartic coupling $\lambda$ is chosen to be negative and small in magnitude, ensuring that the potential develops a local maximum away from the origin --- a necessary condition for the formation of oscillons \cite{Amin_2010, Amin_2012}. The sextic coupling $\sigma > 0$ stabilizes the potential at large field values and sets the depth of the attractive potential well that supports oscillon formation.

Equation \eqref{2.10} is the final action in the Einstein frame, which we will work with in the rest of the paper. Notice that the $R^2$ terms have been translated into the quartic kinetic term $X^2$. The last term is the effective potential in the Einstein frame, $U(\Phi)$. We arrive at non-canonical quartic kinetic terms by moving from the modified gravity action in the Jordan frame to the Einstein frame via a Weyl transformation. 
The time of hyperkination \cite{SanchezLopez:2023ixx}, which can happen immediately following inflation and before kination, is caused by quartic kinetic terms, a distinctive characteristic of the $R^2$ model. 

We can canonicalize the action by introducing a field redefinition given by:
\begin{equation}
    \frac{d\chi}{d\Phi}=\pm \frac{1}{\sqrt{8\alpha V(\Phi)+G(\Phi)}}
    \label{2.11}
\end{equation}
Hence we obtain the more familiar Canonical form of action as :
\begin{align}
     S&=\int d^4x \sqrt{-g}\left[\frac{R}{2}+X+2\alpha\left[8\alpha V(\chi)+G(\chi)\right]\frac{X^2}{2}-U(\chi)\right] \\
     \text{where} \hspace{1cm}
     U(\chi)&=\frac{V(\chi)}{G(\chi)\big[(8\alpha V(\chi)+G(\chi))\big]}
     \label{2.12}
\end{align}
We will mostly use the Non-Canonical Action \eqref{2.10} unless otherwise stated. 

\subsection{Energy-Momentum Tensor}
We can write the action \eqref{2.10} as :
\begin{equation}
    \textbf{S}=\int d^4x \sqrt{-g}\left[\frac{R}{2}+\mathcal{L}_{\Phi}\right]
    \label{2.13}
\end{equation}
where the $\mathcal{L}_{\Phi}$ is the Lagrangian 
\begin{equation}
   \mathcal{L}_{\Phi}= -\frac{\partial_\mu\Phi\partial^\mu\Phi}{2(8\alpha V+G)}+\frac{\alpha (\partial_\mu\Phi\partial^\mu\Phi)^2}{2(8\alpha V+G)}-\frac{V}{G(8\alpha V+G)}
    \label{2.14}
\end{equation}
We consider the flat FLRW metric :
\begin{equation*}
g_{\mu\nu} =
\begin{pmatrix}
-1 & 0 & 0 & 0 \\
0 & a(t)^2 & 0 & 0 \\
0 & 0 & a(t)^2 & 0 \\
0 & 0 & 0 & a(t)^2 
\end{pmatrix}
\end{equation*}
By principle of least action, $\delta\textbf{S}=0$, we get 
\begin{equation}
    \frac{-1}{4}g_{\mu\nu}R+\frac{1}{2}\frac{\delta R}{\delta g_{\mu\nu}}=\frac{1}{2}g_{\mu\nu}\mathcal{L}_{\Phi}-\frac{\delta \mathcal{L}_{\Phi}}{\delta g_{\mu\nu}}
    \label{2.15}
\end{equation}
The RHS of \eqref{2.15} is, by definition, Energy Momentum (Stress-Energy) Tensor $T_{\mu\nu}$. which now takes the form as follows :
\begin{equation}
    T_{\mu\nu}=-2\frac{\delta\mathcal{L}_{\Phi}}{\delta g^{\mu\nu}}+g_{\mu\nu}\mathcal{L}_{\Phi}
\end{equation}
Substituting the Lagrangian and performing the differentiation we get :
\begin{equation}
    T_{\mu\nu}=\frac{1+4\alpha X}{8\alpha V+G}\partial_{\mu}\Phi\partial_{\nu}\Phi+g_{\mu\nu}\left[\frac{X}{(8\alpha V+G)}+X^2\frac{2\alpha}{(8\alpha V+G)}-\frac{V}{G(8\alpha V+G)}\right]
    \label{2.17}
\end{equation}
Now to compute the pressure and Energy density we consider the perfect fluid ansatz for which the Energy-Momentum tensor is given by :
\begin{equation}
    T_{\mu\nu}=(\rho+p)u_{\mu}u_{\nu}+pg_{\mu\nu}
    \label{2.18}
\end{equation}
The invariant square of the 4-velocity of a perfect fluid is normalized to -1, $u_{\mu}u^{\mu}=-1$. For a scalar field acting as a perfect fluid, the 4-velocity is normalized by the gradient of the field :
\begin{equation}
    u_{\mu}=\frac{\partial_{\mu}\Phi}{\sqrt{2X}}
    \label{2.19}
\end{equation}
In our metric signature for a time dependent cosmological background, $\Phi=\Phi(t)$, the kinetic term $X=-\frac{1}{2}g^{00}\dot{\Phi}^2=\frac{1}{2}\dot{\Phi}^2$ is strictly positive, ensuring the term in the square root is real and positive. Hence we get :
\begin{equation}
    u_{\mu}u_{\nu}=\frac{\partial_{\mu}\Phi\partial_{\nu}\Phi}{2X}
    \label{2.20}
\end{equation}
Mapping the terms of \eqref{2.17} and \eqref{2.18} we obtain :
\begin{equation}
    p=\frac{X}{8\alpha V+G}+X^2\frac{2\alpha}{8\alpha V+G}-\frac{V}{G(8\alpha V+G)}
    \label{2.21}
\end{equation}
and 
\begin{equation}
    \rho+p=\frac{8\alpha X^2+2X}{8\alpha V+G}
    \label{2.22}
\end{equation}
Extracting the Energy Density 
\begin{equation}
    \rho=\frac{6\alpha X^2+X}{(8\alpha V+G)}+\frac{V}{G(8\alpha V+G)}
    \label{2.23}
\end{equation}
The Cosmological Equation of State (EOS) parameter is defined as :
\begin{equation}
    \
    \mathcal{\omega}=\frac{p}{\rho} \implies \mathcal{\omega}=\frac{2\alpha G X^2+GX-V}{6\alpha GX^2+GX+V}
    \label{2.24}
\end{equation}
For a homogeneous scalar field :
\begin{equation}
     \mathcal{\omega}=\frac{\alpha G \dot\Phi^4+G\dot\Phi^2-2V}{3\alpha G\dot\Phi^4+G\dot\Phi^2+2V}
     \label{2.26}
\end{equation}
This parameter will transition between the following different regimes depending upon the form of the potential and coupling as well as the value of the scalar field. Broadly they can be described as follows :
\begin{itemize}
    \item \textbf{Vacuum Energy/Slow-Roll limit ($\dot\Phi^2 \xrightarrow{} 0$)}: The kinetic energy of the field is negligible compared to its potential. $\omega \approx -1$. This mimics the nature of the cosmological constant, driving accelerated cosmic expansion.
    \item \textbf{Canonical Kinetic Domination ($\alpha \xrightarrow{} 0$ and $\dot\Phi^2 \gg V$)}: $\omega \approx 1$. This corresponds to a stiff fluid where the speed of sound is equal to the speed of light.
    \item \textbf{Non-Canonical/High-Energy Limit ($\alpha \dot\Phi^4 \gg \dot\Phi^2$)}: This is the ultra-kinetic region where the $X^2$ term dominates and governs the dynamics. Here $\omega \approx \frac{1}{3}$. In this limit, the scalar field behaves like a radiation fluid.
\end{itemize}
In general, if they occur, these regimes appear in the following chronological order: Slow-Roll, Non-Canonical Kinetic Domination (Hyperkination), and finally Canonical Kinetic Domination, before the field settles into oscillations. The slow-roll regime ends when $\epsilon(\Phi) = 1$, marking the end of inflation. Following inflation, a phase of hyperkination driven by the $\alpha X^2$ term can occur if the field velocity is extremely large ($\alpha X^2 \gg X$). However, whether these kinetic-dominated stages appear is highly dependent on the choice of model parameters. For our specific parameters, the kinetic energy remains relatively small ($X^2 \ll 1$) throughout the post-inflationary evolution. Because the threshold $\alpha X^2 \sim \mathcal{O}(1)$ is never reached, the non-canonical corrections are strongly suppressed, meaning the hyperkination stage is entirely bypassed. Furthermore, the field does not experience a distinct canonical kinetic domination phase (where $X \gg V$) because it falls immediately into the potential well and begins oscillating, rapidly establishing an equipartition between kinetic and potential energy. This explains why a separate stage of kinetic domination is not visibly present in our lattice results; it is subsumed by the immediate onset of the oscillatory preheating phase. Since $X^2 \ll 1$ is satisfied throughout, the post-inflationary dynamics are well-approximated by the canonical Einstein-frame action, justifying the numerical approach of Section~\ref{sec:numerical}.

\section{Linear Stability Analysis and Floquet Theory}
\label{sec: linear stability}
Tachyonic resonance plays an important role in the growth of perturbations during inflationary preheating, leading to the necessity of investigating the instability bands in the momentum space of comoving $k$-modes and studying their trajectory as they pass through these instability bands. Floquet theory can be efficiently used to perform a linear stability analysis of the inflaton fluctuations $\delta\chi_k$. We identify the instability bands by applying Floquet theory. The detailed mathematical approach to Floquet theory can be found in the appendices.
\subsection{Linear Perturbation Theory for Fluctuations}
The Klein-Gordon equation, ignoring the $X^2$ term in the action (the reason is argued later in Section \ref{sec:numerical}), for the canonical field $\chi$ in the Einstein frame, governing the full non-linear dynamics of the background homogeneous field and the fluctuations, is given by:
\begin{equation}
  \ddot{\chi}+3H\dot{\chi}-\frac{1}{a^2}\nabla^2\chi+\frac{dU}{d\chi}=0
  \label{3.1}
\end{equation}
We can split the field as $\chi(t,x)=\chi_0(t)+\delta\chi(t,x)$. The homogeneous background will satisfy the spatially averaged equation :
\begin{equation}
    \ddot{\chi}_0+3H\dot{\chi}_0+\frac{dU}{d\chi}|_{\chi_0}=0
    \label{3.2}
\end{equation}
and the perturbations upto linear order in $\delta\chi$ satisfy :
\begin{equation}
    \delta\ddot{\chi}+3H\delta\dot{\chi}-\frac{\nabla^2\delta\chi}{a^2}+\frac{d^2U}{d\chi^2}|_{\chi_0}\delta\chi=0
    \label{3.3}
\end{equation}
This is the linearized perturbation equation in real space.
Now performing Fourier decomposition and transforming to Momentum space using :
\begin{equation}
    \delta\chi(t,x)=\int\frac{d^3k}{(2\pi)^3}\delta\chi_k(t)e^{i\textbf{k}\cdot \textbf{x}}
    \label{3.4}
\end{equation}
We obtain the $k$ space version of \eqref{3.3} :
\begin{equation}
    \delta\ddot{\chi}_k+3H\delta\dot{\chi}_k+\left[\frac{k^2}{a^2}+m^2_{eff}(t)\right]\delta\chi_k=0
    \label{3.5}
\end{equation}
where the time dependent \textbf{Effective Mass Squared} term is given by :
\begin{equation}
    m^2_{eff}(t)=\frac{d^2U}{d\chi^2}|_{\chi_0(t)}
    \label{3.6}
\end{equation}

\noindent \textbf{Initial Conditions (Bunch Davies Vaccum)} : The vaccum fluctuations that seed the subsequent instability takes this form -
\begin{equation}
    \delta\chi_k(0)=\frac{1}{a\sqrt{2k}} \hspace{0.5cm} \text{and} \hspace{0.5cm} \delta\dot{\chi}_k(0)=\frac{-ik}{a}\delta\chi_k(0)
    \label{3.7}
\end{equation}

\noindent \textbf{Flat Space limit} : For our values of model parameters we have $\frac{H_{end}}{m} \approx 8.96 \times10^{-4}<<1$ so we are almost in the Minkowski space so setting $a=1$ and neglecting the Hubble friction, $3H$, term, the equation \eqref{3.5} simplified to :
\begin{equation}
    \delta\ddot{\chi}_k+\omega^2_k(t)\delta\chi_k=0
    \label{3.8}
\end{equation}
with $\omega^2_k(t)=k^2+\frac{d^2U}{d\chi^2}|_{\chi_0(t)}$ where we have neglected metric perturbations and this type of equation is known as \textit{Hill's Differential Equation} \cite{McLachlan1947}, which according to Floquet's Theorem have solutions of the form :
\begin{equation}
    \delta\chi_k(t)=e^{\mu_k}\mathcal{P}_{(+)}(t)+e^{-\mu_k}\mathcal{P}_{(-)}(t)
    \label{3.10}
\end{equation}
$\mu_k$ in the exponents are called Floquet exponents and they govern how the $k$ modes vary with time. The functions $\mathcal{P}_{(\pm)}(t)$ are functions with period same as that of the oscillation of the background field $\chi_0(t)$. There are effectively three regimes which can be identified based on the nature of Real part of $\mu_k$, $\mathcal{R}(\mu_k)$ :
\begin{itemize}
    \item $\mathcal{R}(\mu_k)>0$ : Exponential growth, Unstable.
    \item $\mathcal{R}(\mu_k)=0$, $\mathcal{I}m(\mu_k)\ne0$ : Oscillatory, Stable
    \item $\mathcal{R}(\mu_k)=0$, $\mathcal{I}m(\mu_k)=0$ : Secular Growth, Resonance boundary
\end{itemize}
\textbf{Effective Mass Squared $m^2_{eff}$} : We make use of the chain rule to find $\frac{d^2U}{d\chi^2}$, which ultimately gives :
\begin{equation}
    m^2_{eff}=G(\Phi)\frac{d^2U}{d\Phi^2}+\frac{G'(\Phi)}{2}\frac{dU}{d\Phi}
    \label{3.11}
\end{equation}
\textbf{Tachyonic Region} : The instability occurs when $m^2_{eff}<0$. From equation \eqref{3.5} and \eqref{3.6} we can imply that (which also follows from \cite{Piani_2025}), the maximum unstable wavenumber at a given instant is :
\begin{equation}
    k_{tach}=a\sqrt{|m^2_{eff}(\chi_0(t))|}
    \label{3.12}
\end{equation}
where $k_{\rm tach}$ is the comoving tachyonic momentum, related to the physical momentum by $k_{\rm phys} = k_{\rm tach}/a$. Only modes with $k<k_{tach}$ are unstable at that instant.

\noindent \textbf{Background Solutions and Oscillation Period} : In Minkowski limit \eqref{3.2} reduces to that of a Simple Harmonic Oscillator, and its first integral gives the energy conservation equation :
\begin{equation}
    \frac{1}{2}\dot{\chi}^2_0+U(\chi_0)=E=\text{Constant}
    \label{3.13}
\end{equation}
Now we assume the end of inflation and start of simulation as $t=0$, so at this moment $\chi_0(t=0)=\chi_{end}$ and $\dot{\chi}_0(t=0)=0$.

The second initial condition can be justified by the following reasoning : Think of this system analogous to that of pendulum where $\chi_{end}$ represents the value at the point of maximum displacement of the bob (Inflaton in our case). So when we release the pendulum at $t=0$, its initial velocity is zero.
So the initial total energy of the system is $E=U(\chi_{end})$ and this remains constant throughout as there is no damping term. So we can write \eqref{3.13} as 
\begin{equation}
     \frac{1}{2}\dot{\chi}^2_0+U(\chi_0)=U(\chi_{end})
     \label{3.14}
\end{equation}
so the background solutions at any general time $t$ are given by :
\begin{equation}
    \dot{\chi}_0=\pm\sqrt{2\left[U(\chi_{end})-U(\chi_0)\right]}
\end{equation}
For symmetric potential as in our case the time period of oscillation $T$ (complete period from positive to negative end) is obtained by solving the integral :
\begin{equation}
    T(\chi_{end})=4\int_{0}^{\chi_{end}}\frac{d\chi}{\sqrt{2\left[U(\chi_{end})-U(\chi_0)\right]}}
    \label{3.16}
\end{equation}
\subsection{Floquet Theory}
We can now solve \eqref{3.8} by writing it as a system of coupled first order differential equation in matrix form :
\begin{equation}
\frac{d}{dt}
\begin{pmatrix}
\delta \chi_k \\
\delta \pi_k
\end{pmatrix}
=
\mathbf{U}(t)
\begin{pmatrix}
\delta \chi_k \\
\delta \pi_k
\end{pmatrix},
\label{3.17}
\end{equation}
where
\begin{equation}
\delta \pi_k \equiv \dot{\delta \chi}_k,
\label{3.18}
\end{equation}
and
\begin{equation}
\mathbf{U}(t) =
\begin{pmatrix}
0 & 1 \\
- k^2 - m_{\mathrm{eff}}^2(t) & 0
\end{pmatrix}.
\label{3.19}
\end{equation}

Now we define two orthogonal initial conditions :
\begin{equation}
    \{\delta\chi_k^{(1)}(0)=1, \hspace{0.1cm} \delta\pi_{k}^{(1)}(0)=0 \} \hspace{0.3cm} \text{and} \hspace{0.3cm} \{\delta\chi_k^{(2)}(0)=0, \hspace{0.1cm} \delta\pi_{k}^{(2)}(0)=1 \}
    \label{3.20}
\end{equation}
To solve this system we evolve it from $t=0$ to $T$ using a method known as \textbf{Matrix Monodromy Method}. In this section we only provide the final expressions for the Floquet exponents $\mu_k$, the detailed derivation can be found in the Appendices \ref{Appendix A}.
\begin{equation}
\mathcal{R}\!\left(\mu_k^{\pm}\right)
= \frac{1}{T}
\ln\left|
\frac{\delta\chi_k^{(1)}(T) + \delta\pi_k^{(2)}(T)}{2}
\pm
\sqrt{
    \frac{\left[\delta\chi_k^{(1)}(T) - \delta\pi_k^{(2)}(T)\right]^2}{4}
    + \delta\chi_k^{(2)}(T)\,\delta\pi_k^{(1)}(T)
}
\right|
\label{3.21}
\end{equation}
where the superscripts $(1)$ and $(2)$ correspond to the two solutions evolved from the orthogonal initial conditions \eqref{3.20} and all quantities are evaluated at $t=T$. The physical Floquet Exponent is given by :
\begin{equation}
\mathcal{R}(\mu_k)
= \max\!\left[
    \mathcal{R}\!\left(\mu_k^{+}\right),\;
    \mathcal{R}\!\left(\mu_k^{-}\right),\;
    0
\right]
\end{equation}
The plot for the variation of Floquet Exponents and the Floquet chart is shown below.

\begin{figure}[htbp]
    \centering
    \includegraphics[width=0.95\textwidth]{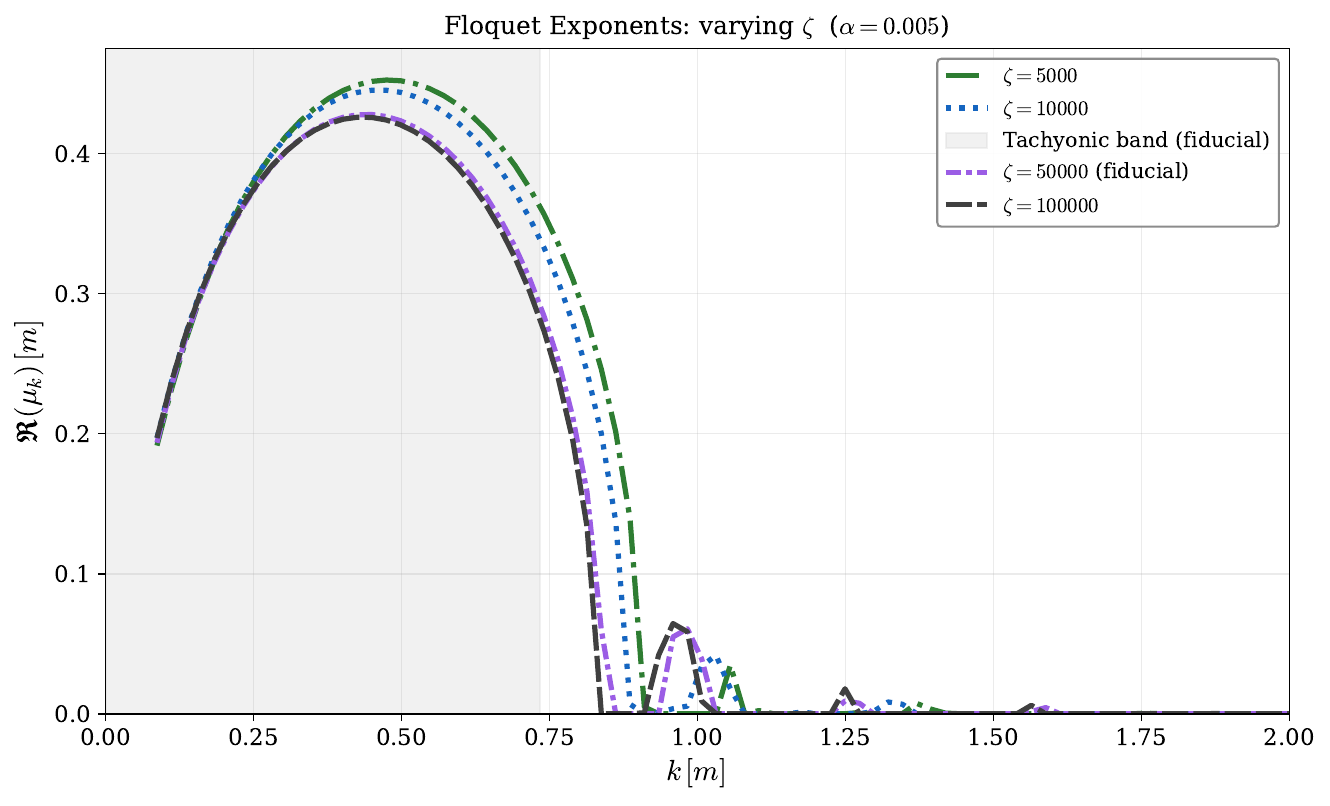}
    \caption{Variation of the real part of the Floquet 
    exponents $\mathfrak{R}(\mu_k)/m$ vs $k/m$ for varying 
    $\zeta \in \{5000, 10000, 50000, 100000\}$ with 
    $\alpha = 0.005$ fixed. The gray shaded region 
    representing the Tachyonic Resonance band.}
    \label{fig:floquetexponents}
\end{figure}

The broad tachyonic instability persists across all parameter values with peak Floquet exponents $\mathfrak{R}(\mu_k)/m \in [0.43, 0.45]$, confirming robust oscillon formation throughout the explored parameter space. Smaller values of $\zeta$ exhibit a slightly wider and stronger instability band, while larger $\zeta$ leads to mild suppression. The narrow parametric resonance peaks at $k/m \approx 0.95$ and $k/m \approx 1.25$ persist across all values.

We also investigated the effect of varying $\alpha \in \{0.001, 0.005, 0.01\}$ while keeping $\zeta = 50000$ fixed. The resulting Floquet spectra are found to be virtually identical across all three values, with peak exponents $\mathfrak{R}(\mu_k)/m \approx 0.428$ and tachyonic band boundary $k_{\rm tach}/m \approx 0.734$ unchanged. This insensitivity arises because the post-inflationary potential is dominated by the non-minimal coupling $\zeta$, while the Palatini $R^2$ correction enters through $8\alpha V(\Phi)$, which is negligibly small compared to $G(\Phi) = 1 + \zeta\Phi^2$ for field values near the minimum. Consequently, the Floquet spectrum and oscillon formation are effectively independent of $\alpha$ in the physically motivated range, and we do not show a separate figure for this case.

A perturbation $\delta\chi_k$ with wavenumber $k$ grows in amplitude as $\delta\chi_k \propto e^{\mathcal{R}(\mu_k)t}$. The massive blue ``hump'' in the plot grows rapidly in the tachyonic instability region ($0 < k < 0.73$) and decreases as it comes out of the tachyonic band, thus representing that the perturbations grow exponentially by a factor of $\approx e^{0.43}$ and attain their maximum amplitude in that region; this is where the homogeneity of the field breaks and oscillon formation begins. The tiny blue spikes are due to a phenomenon known as \textbf{Parametric Resonance} \cite{Zlatev_1998}. They occur in the region where $m^2_{\rm eff} > 0$, but the modes get excited by the oscillating background. As is clearly visible, the kick in perturbation due to this is very tiny as compared to those produced by the tachyonic resonance.

\begin{figure}[htbp]
    \centering
    \includegraphics[width=0.85\textwidth]{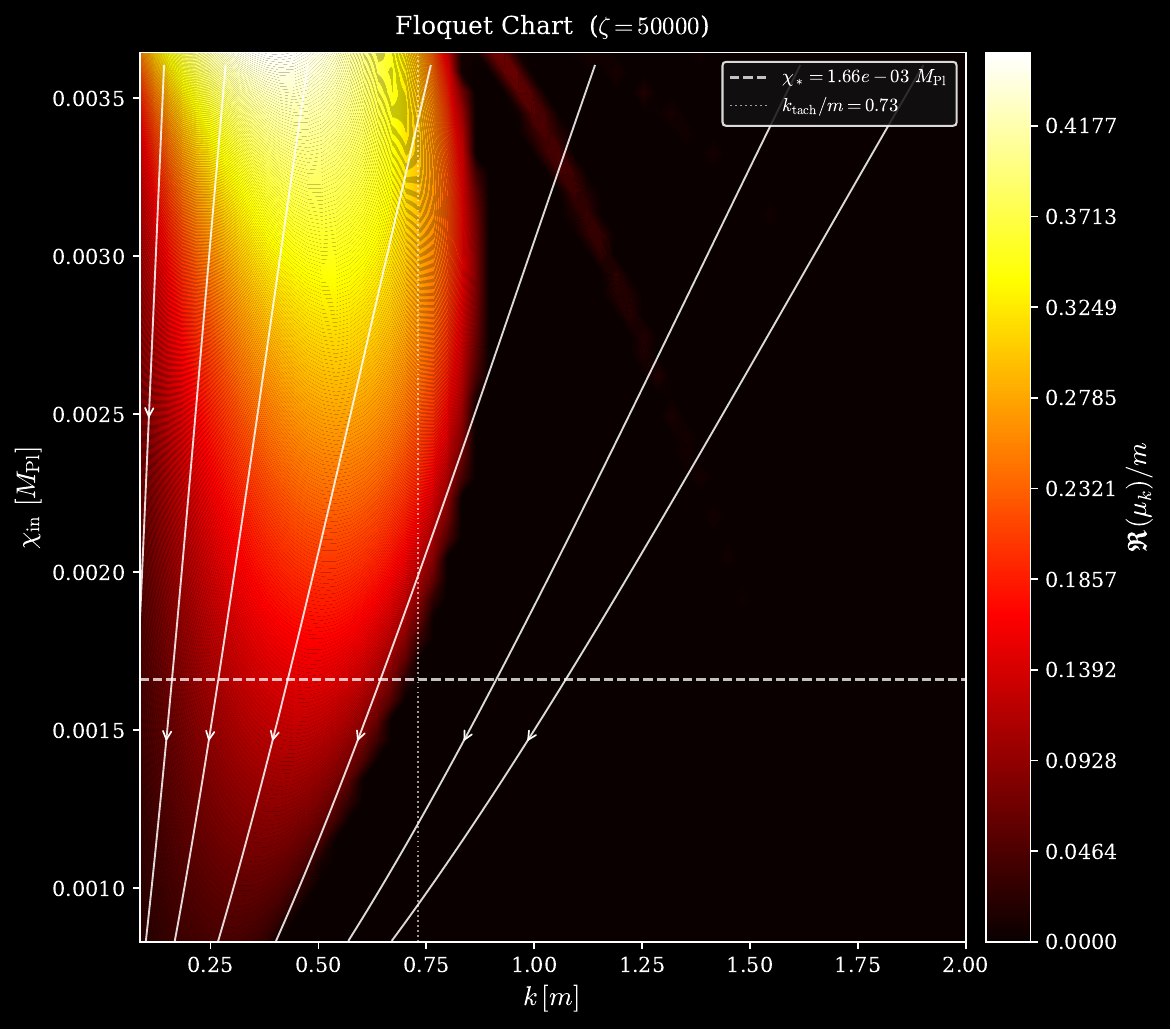}
    \caption{Floquet chart with the color bar representing 
    the real part of the Floquet exponent.}
    \label{fig:floquetchart}
\end{figure}

The Floquet chart in Figure~\ref{fig:floquetchart} shows the existence of a broad instability band for $k \lesssim 0.75$, where $\mathcal{R}(\mu_k)|_{\rm max} \approx 0.42$, and a narrow band represented in red for $1.00 \lesssim k \lesssim 1.5$. The white flow lines indicate how the physical $k$-modes pass through different resonance bands as they evolve due to the expansion of the universe. One thing to note is that these specific values of the Floquet exponents and the tachyonic wavenumber are only for our particular choice of model parameters and can change if we choose a different set of parameters. This explains how the power spectrum should behave: for wavenumbers $k \lesssim 0.8$, there should be a sharp enhancement in the perturbations at relatively early times.

\section{Oscillon Dynamics: Numerical Approach}
\label{sec:numerical}
We numerically replicate the physics of the early universe and examine the conditions and requirements for the emergence of oscillons. To extract the actual oscillon dynamics and study their formation, we perform what is known as non-perturbative $3+1$ classical lattice simulations using the publicly available package \textit{Cosmo}$\mathcal{L}$\textit{attice} \cite{Figueroa_2023}.\newpage One is free to choose among other similar software packages --- \textit{LatticeEasy} \cite{Felder_2008}, DEFROST \cite{Frolov_2008}, and HLATTICE \cite{Huang_2011}.
\subsection{Field and Potential Redefinition}
Equation \eqref{2.10} reveals that the Einstein-frame action contains a non-canonical quartic kinetic term proportional to $X^2$. Physically, this higher-derivative term drives the phase of hyperkination \cite{Kuralkar:2025hoz}. However, the formation and dynamics of oscillons occur during the post-inflationary epoch, where the scalar field oscillates near the minimum of its potential. In this low-energy regime, the kinetic energy of the field is small ($X \ll 1$), rendering higher-order kinetic corrections dynamically strongly suppressed. Moreover, the coefficient of the $X^2$ term is dominated by $\alpha$, which we assume to be $10^{-3} (\ll 1)$, safe enough to neglect. Thus, we can rigorously approximate the system using the standard canonical Einstein-Hilbert action without loss of physical accuracy, allowing the model to be securely evolved using the framework that is most common in all simulation software packages.
Let
\begin{align}
    F(\Phi)&=8\alpha V(\Phi)+G(\Phi)\\
    F(\Phi)&=1+A\Phi^2+B\Phi^4+C\Phi^6
    \label{4.1}
\end{align}
with $A=4\alpha m^2+\zeta$, $B=-2\alpha \lambda$ and $C=\frac{4\alpha \sigma}{3 m^2}$. Now following \ref{2.11} we can write : 
\begin{equation}
     \frac{d\chi}{d\Phi}=\frac{1}{\sqrt{1+A\Phi^2+B\Phi^4+C\Phi^6}}
\implies \frac{d\chi}{d\Phi}=\frac{1}{\sqrt{1+A\Phi^2}}\left[1-\frac{B\Phi^4+C\Phi^6}{2(1+A\Phi^2)}\right]
\label{4.2}
\end{equation}
The term in the square bracket is sufficiently less than 1 considering the values of the potential and hence can be safely neglected, so the integral becomes :
\begin{align}
    \int d\chi&=\int \frac{1}{\sqrt{1+A\Phi^2}}d\Phi \\
    \chi&=\frac{1}{\sqrt{A}}\sinh^{-1}({\sqrt{A}\Phi)} \implies\Phi=\frac{1}{\sqrt{A}}\sinh{\sqrt{A}\chi}
    \label{4.4}
\end{align}
\textbf{Crossover Scale} : The change from canonical to non-canonical dynamics in the Einstein frame is controlled by a field-dependent prefactor that modifies the scalar field's kinetic term. This prefactor is defined as $F(\Phi) = 8\alpha V(\Phi) + G(\Phi)=1+A\Phi^2+B\Phi^4+C\Phi^6$. In order for the theory to naturally reduce to standard General Relativity with a canonical kinetic term, we must recover the limit $F(\Phi) \simeq 1$, which requires $A\Phi^2<<1$. The crossover scale, $\Phi_c$, is defined as the threshold above which the effects of modified gravity term $R^2$ become dynamically dominant. When the field-dependent kinetic corrections compete with the canonical constant term, this physical change takes place, namely when $A\Phi_c^2 \simeq 1$. Therefore, the crossover scale is analytically defined as:
\begin{equation}
\Phi_c = \frac{1}{\sqrt{A}} = \frac{1}{\sqrt{4\alpha m^2 + \zeta}} \simeq \frac{1}{\sqrt{A}}\approx 0.00447 M_{\text{pl}}
\label{4.6}
\end{equation}
We can summarize the field values as :
\begin{equation}
\Phi \simeq 
\begin{cases} 
\chi  & \chi < \chi_c \\[8pt]
\frac{1}{2\sqrt{A}}e^{\sqrt{A}\chi} & \chi \gg \chi_c
\end{cases}
\label{4.7}
\end{equation}
Similarly we examine the Effective potential behavior in both regimes:
\begin{equation}
U(\chi) = \begin{cases} 
\frac{m^2 \chi^2}{2} - \frac{\lambda \chi^4}{4} + \frac{\sigma \chi^6}{6m^2} & \chi < \chi_c \\[8pt]
\frac{2m^2}{\zeta} e^{-2\sqrt{A}\chi} - \frac{\lambda}{4\zeta A} + \frac{\sigma}{24m^2 \zeta A^2} e^{2\sqrt{A}\chi} & \chi \gg \chi_c
\end{cases}
\label{4.8}
\end{equation}
\subsection{Post Inflationary Phase : Preheating}
Following the conclusion of inflation, the inflaton field can be effectively characterized as a homogeneous condensate, with the bulk of its energy concentrated in its zero momentum mode ($k=0$). Its dynamics is governed by the Klein Gordon Equation in a flat FLRW spacetime :
\begin{equation}
    \ddot{\chi}+3H\dot{\chi}-\frac{1}{a^2}\nabla^2\chi+U'(\chi)=0
    \label{4.9}
\end{equation}
where $a$ is the scale factor and $H\equiv\frac{\dot{a}}{a}$ is the Hubble rate which is determined by the Friedmann Equations 
\begin{equation}
   3H^2 = \frac{1}{2} \dot{\chi}^2 + \frac{1}{2a^2} (\nabla \chi)^2 + U(\chi) \hspace{2cm} \dot{H} = -\frac{1}{2} \dot{\chi}^2 - \frac{1}{6a^2} (\nabla \chi)^2
   \label{4.10}
\end{equation}
Overdots represent the derivative with respect to cosmic time, $t$. In a purely homogeneous universe, assuming no particle production, the zero mode oscillates about the potential minimum as its amplitude drops adiabatically due to Hubble friction, where the energy density scales as $\rho \propto a^{-3(1+\bar{\omega})}$ \cite{PhysRevD.28.1243}, giving a field amplitude $\Phi \propto \rho^{1/2} \propto a^{-3(1+\bar{\omega})/2}$, consistent with $\Phi \propto a^{-3/2}$ for $\bar{\omega} = 0$. where $\bar{\omega}$ is the time averaged equation of state :
\begin{equation}
    \bar{\omega}=\frac{\langle p \rangle}{\langle \rho \rangle} \simeq\frac{n-1}{n+1}
    \label{4.11}
\end{equation}
where $n$ is determined from the potential behavior around its minimum, $\Phi^{2n}$. Hence if the potential acts as quadratic around its minimum i.e $n=1$ then we have $\bar{\omega}=0$ and the universe is matter dominated with the amplitude decreasing as $\Phi \propto a^{-3/2}$. For a potential behaving as quartic around the minimum, $n=2$, we have radiation dominated epoch with $\bar{\omega}=\frac{1}{3}$.

The non-linear preheating dynamics, which is a significant feature of our Palatini modified gravity model, can substantially influence post-inflationary evolution, despite the absence of direct interactions involving other matter fields.
As a consequence, the homogeneous zero mode exhibits substantial instability. Rather than undergoing a gradual and adiabatic decline due to Hubble friction, the condensate is compelled to abruptly release its energy into designated non-zero momentum bands ($k \neq 0$). 

The early phases of particle production can be effectively characterized by a linearized analysis. Consequently, we shall express the inflaton as a combination of the background and its fluctuations for the redefined field, $\chi(x,t)= \bar{\chi}(t)+\delta\chi(x,t)$,then analyze the evolution of the latter in Fourier (momentum) space. At leading order, the Fourier modes $\delta\chi_k$ satisfy the equation :
\begin{equation}
\ddot{\delta \chi}_k + 3H \dot{\delta \chi}_k + \left( \frac{k^2}{a^2} + U_{,\chi\chi} \right) \delta \chi_k = 0
\label{4.12}
\end{equation}
where $k$ is the associated wave-vector. 

For field values greater than the inflection point, $\chi_i$, the second derivative of the potential turns negative, initiating a \textbf{tachyonic} instability \cite{Felder_2001} in \eqref{4.12}. The field sits at a local maximum of its potential energy, making it highly unstable to small perturbations. The effective square frequency of this damped oscillator system becomes non-positive definite as the inflaton field accelerates towards the potential minimum from beyond the inflection point. Mathematically, the field behaves like it has an imaginary mass. This leads to the exponential amplification of long-wavelength fluctuations with $\textbf{k}^2 < a^2|U_{,\chi\chi}|$. It induces spontaneous symmetry breakdown and may lead to ``tachyonic preheating'', a phase where there is rapid conversion of potential energy into scalar waves. The exponential growth of fluctuations can be interpreted as the growth of the occupation number of particles with $\textbf{k}^2 < a^2|U_{,\chi\chi}|$. 

If the resultant fluctuations are not effectively suppressed by the universe's expansion, they will ultimately influence the homogeneous dynamics, culminating in the fragmentation of the inflaton condensate into a significantly inhomogeneous state characterized by substantial overdensities, which subsequently leads to the production of oscillons.
Due to the concurrent exponential growth of an entire band of $k$-modes, the previously homogeneous field is swiftly fragmented into substantial, high-contrast spatial fluctuations. This violent disruption of the vacuum is termed \textbf{spinodal decomposition} \cite{PhysRevD.62.023520}.
The maximum momentum resulting from the tachyonic instability is determined by the minimum of the second derivative of the potential \eqref{4.8}, which turns out to be a parametrically fixed value:
\begin{equation}
    \chi_{min} = \frac{1}{4\sqrt{A}} \ln \left( \frac{48 m^4 \xi A^2}{\zeta \sigma} \right)
    \label{4.13}
\end{equation}
such that 
\begin{equation}
    \frac{|k_{\text{max}}|}{a} = \sqrt{|U''(\chi_{min})|}= 2 \left( \frac{\sigma}{3\zeta^2} \right)^{1/4}
    \label{4.14}
\end{equation}

\subsection{\textit{Cosmo}$\mathcal{L}$\textit{attice} Setup}
In this section we begin with fixing the $3d$ simulation parameters, potential and initial conditions that our best suited for our model.
Lattice simulations generally work with dimensionless variables with appropriate re-scalings, so we first do the necessary transformations from physical to program variables : 
\begin{equation}
    \tilde{\chi} \equiv \frac{\chi}{f_{*}} \hspace{1cm} \tilde{\eta}\equiv a^{-\alpha_1}\omega_{*}\tau \hspace{1cm} \tilde{x}^i\equiv\omega_{*}x^i \hspace{0.5cm }\text{(i=1,2,3 represent spatial dimensions)}
    \label{4.15}
\end{equation}
where $\chi$ is the scalar field, $\tilde{\eta}$ is the program time and $f_{*}$ and $\omega_{*}$ are constants with dimensions of energy in $GeV$. We choose $\omega_{*}=m = 12.2 \times10^{-12} GeV$ and $\alpha_1=3\frac{p-2}{p+2}$, where p is the power of the dominant term in the potential for our case $p=4$ which makes $\alpha_1=1$ and the code uses the conformal time $d\tilde{\eta}\propto \frac{d\tau}{a}$.
The parameter $\alpha_1 = 1$ corresponds to conformal time 
$d\tilde{\eta} \propto a^{-1}dt$, chosen according to the 
CosmoLattice prescription \cite{Figueroa_2023}:
\begin{equation}
    \alpha_1 = \frac{3(p-2)}{p+2}
\end{equation}
which guarantees that the oscillation frequency of the 
program field variable remains approximately constant when 
expressed in $\alpha_1$-time, allowing each physical 
oscillation to be resolved with similar accuracy throughout 
the simulation.

We also note that while the leading term in the Taylor expansion 
of $U(\chi)$ around the minimum is quadratic ($p = 2$, 
suggesting $\alpha_1 = 0$, i.e.\ cosmic time), the dynamics 
during the initial fragmentation phase is dominated by the 
quartic structure of the potential ($p = 4$, giving 
$\alpha_1 = 1$) when the field is traversing the potential 
barrier between the inflationary plateau and the minimum. 
Following the CosmoLattice recommendation \cite{Figueroa_2023}, 
we use $\alpha_1 = 1$ (conformal time) during this phase. 
Once the field settles into the quadratic regime near the 
minimum, the choice of time variable does not affect the 
physical results.

The potential we input in the code is obtained using the relation \eqref{4.4} and for the current values of our model parameters the effective Einstein frame potential $U(\Phi)$ given by \eqref{2.12} can be approximated as 
\begin{equation}
    U(\Phi(\chi))=\frac{V(\Phi(\chi))}{[G(\Phi(\chi))]^2}
    \label{4.16}
\end{equation}
Now substituting $\Phi(\chi)=\ \frac{M_{pl}}{\sqrt{\xi}} \sinh\left( \frac{\sqrt{\zeta} \chi}{M_{pl}} \right)$ we get :
\begin{equation}
  U(\chi) = \left[ \frac{m^2 M_{pl}^2}{2\zeta} \right] \tanh^2(s)\text{sech}^2(s) - \left[ \frac{\lambda M_{pl}^4}{4\zeta^2} \right] \tanh^4(s) + \left[ \frac{\sigma M_{pl}^6}{6 m^2 \zeta^3} \right] \sinh^2(s)\tanh^4(s)
  \label{4.17}
\end{equation}
with $s=\frac{\sqrt{\zeta} \chi}{M_{pl}}$. We can further convert it into program potential by using $\chi=\tilde{\chi}f_{*}$.
The Following parameters define the geometry of the lattice :
\begin{itemize}
    \item \textbf{N} : Number of lattice points along each edge of the cube. Total $N^3$ points. We consider $N=288$.
    \item\textbf{L} : The Physical size of the box, modes with $\lambda>L$ cannot be represented.
    \item\textbf{dx} : Spacing between lattice points $dx=L/N$.
    \item\textbf{kcutoff} : Momentum cutoff value for the initial quantum fluctuations. Only modes with $k<kcutoff$ receive the \textbf{Bunch Davies} noise \cite{launay2025bunchdaviesinitialconditionsnonperturbative}. 
    \item\textbf{kIR} : co-moving wavenumber of the longest wave that fits inside the simulation box. $kIR=\frac{2\pi}{L}$. We need to specify either $L$ or $kIR$, the other is calculated by the code internally.
\end{itemize} 
We run the simulation for $\tilde{\eta}=500$ with a timestep of $dt=0.05$ and box size $L=72.4  M^{-1}$ 

\vspace{1em}
\noindent \textbf{Initial Conditions} 
\vspace{0.5em}

The simulation starts when inflation ends, so we need to provide the initial field amplitude, which is the value of the canonical field at the end of inflation. This is determined by the condition $\epsilon(\Phi) = 1$, where the potential slow-roll parameter is given by:
\begin{equation}
    \epsilon(\Phi) = \frac{1}{2}\left[8\alpha V(\Phi) + G(\Phi)\right] \cdot \left[\frac{1}{U(\Phi)}\frac{dU(\Phi)}{d\Phi}\right]^2
    \label{4.18}
\end{equation}
\begin{equation}
    \eta(\Phi) = k(\Phi)\left[\frac{U''(\Phi)}{U(\Phi)} + \frac{1}{2}\left(\frac{U'(\Phi)}{U(\Phi)}\right)\left(\frac{k'(\Phi)}{k(\Phi)}\right)\right]
    \label{4.20}
\end{equation}
where $k(\Phi) = 8\alpha V(\Phi) + G(\Phi)$. At the end of inflation $\epsilon(\Phi) = 1$, giving:
\begin{equation}
    \Phi_{\rm end} = 4.3245 \times 10^{-3}\,M_{\rm pl} 
    \implies 
    \chi_{\rm end} = 3.8364 \times 10^{-3}\,M_{\rm pl}
    \label{4.21}
\end{equation}
The Hubble rate at the end of inflation is computed from the Friedmann equation:
\begin{equation}
    H^2 = \frac{\rho}{3M_{\rm pl}^2}
    \label{4.22}
\end{equation}
Since kinetic and potential energies are comparable at the end of inflation, the total energy density is:
\begin{equation}
    \rho = \frac{1}{2}\dot{\chi}^2 + U(\chi)
    \label{4.23}
\end{equation}
The Hubble rate at the end of inflation is then:
\begin{equation}
    H_{\rm end} = \sqrt{\frac{U(\chi_{\rm end})}{3M_{\rm pl}^2}} = 4.480 \times 10^{-9}\,M_{\rm pl}
    \label{4.24}
\end{equation}
where in Eq.~(4.24) we have used the slow-roll approximation $\rho \approx U$ which holds at the onset of the post-inflationary phase.

The initial field velocity $\dot{\chi}_{\rm end}$ is determined not from the slow-roll Klein-Gordon equation, which breaks down precisely at $\epsilon = 1$, but from the exact condition at the end of inflation. Setting the Hubble slow-roll parameter $\epsilon_H \equiv -\dot{H}/H^2 = 1$ gives $w = -1/3$ via:
\begin{equation}
    \epsilon_H = \frac{3}{2}(1 + w) = 1 
    \implies 
    w = -\frac{1}{3}
    \label{4.25}
\end{equation}
Writing the equation of state as $w = p/\rho$ with $p = \frac{1}{2}\dot{\chi}^2 - U$ and $\rho = \frac{1}{2}\dot{\chi}^2 + U$:
\begin{equation}
    -\frac{1}{3} = \frac{\frac{1}{2}\dot{\chi}^2 - U}{\frac{1}{2}\dot{\chi}^2 + U} 
    \implies 
    \dot{\chi}^2_{\rm end} = U(\chi_{\rm end})
    \label{4.26}
\end{equation}
Using $U(\chi_{\rm end}) = 3H_{\rm end}^2$ from Eq.~(4.24):
\begin{equation}
    \dot{\chi}_{\rm end} = \sqrt{3}\, H_{\rm end} = \sqrt{3} \times 4.480 \times 10^{-9} = 7.759 \times 10^{-9}\, M_{\rm pl}^2
    \label{4.27}
\end{equation}
The negative sign is chosen since the field is rolling toward the potential minimum.

The scale factor is evolved using the second Friedmann equation:
\begin{equation}
    \frac{\ddot{a}}{a} = \frac{1}{3}
    \langle -\dot{\chi}^2 + U \rangle
    \label{4.28}
\end{equation}
The first Friedmann equation
\begin{equation}
    H^2 = \frac{1}{3}\left\langle\frac{1}{2}\dot{\chi}^2 
    + \frac{1}{2a^2}(\nabla\chi)^2 + U\right\rangle
    \label{4.29}
\end{equation}
is used to monitor energy conservation. Here the second term is called the ``gradient energy'' and the angular brackets $\langle\ldots\rangle$ denote the spatial averaging over the simulation volume. The equation of state parameter is computed by:
\begin{equation}
    \bar{\omega} = \frac{E_K - \frac{1}{3}E_G - E_V}
    {E_K + E_G + E_V}
    \label{4.30}
\end{equation}
where $E_K$, $E_G$, $E_V$ represent the kinetic, gradient and potential energy respectively.

One important thing to note is that, in reality, oscillons are massive objects and gravitationally attract each other. Over long time scales, they may merge and cluster together to form large structures and sometimes even primordial black holes ($\mathrm{PBHs}$), as studied in \cite{Cotner_2018}. But by using \textit{Cosmo}$\mathcal{L}$\textit{attice}, we cannot study the non-linearities due to gravitational clustering and merging of these objects \cite{Mahbub_2023} because, like many other lattice simulation software packages, it does not take into account the Bardeen potential \cite{2021MNRAS.504.5612D} that can affect the metric perturbations.
We now present our simulation results and discuss their physical significance.

\subsection{Field Dynamics}
\subsubsection{Evolution of Mean Field Values}
The typical growth of perturbations is characterized by the root-mean-square (RMS) value of the inflaton field:
\begin{equation}
    \mathrm{rms}(\chi) = \sqrt{\langle\chi^2\rangle - 
    \langle\chi\rangle^2}
    \label{eq:4.31}
\end{equation}

\begin{figure}[H]
    \centering
    \makebox[\textwidth][c]{%
        \includegraphics[width=1.1\textwidth]{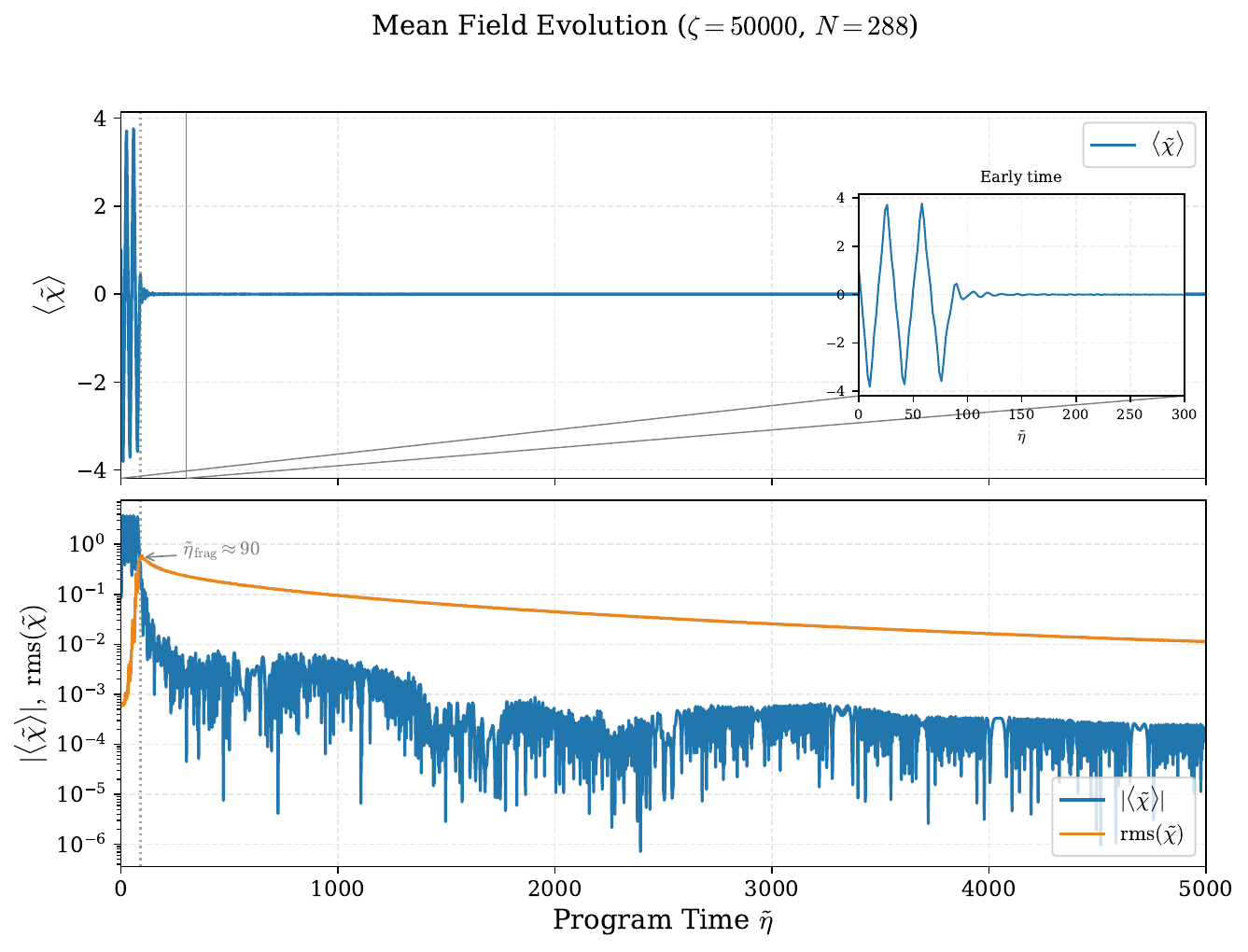}}
    \caption{The evolution of average field value vs 
    simulation time. The top panel shows the backreaction 
    and fragmentation phase taking place at 
    $\tilde{\eta} \approx 90$. The bottom panel shows 
    $|\langle\tilde{\chi}\rangle|$ (blue curve) and 
    $\mathrm{rms}(\tilde{\chi}) \equiv 
    \sqrt{\langle\tilde{\chi}^2\rangle - 
    \langle\tilde{\chi}\rangle^2}$ (orange curve), 
    characterizing the growth of field fluctuations.}
    \label{fig:meanfield}
\end{figure}

\FloatBarrier

\subsubsection{Power Spectrum}

The unstable Tachyonic band in Fig.~\ref{fig:floquetchart} 
and the claim we made at the very end of 
section~\ref{sec: linear stability} can be verified by the 
Two-Point Correlation function in Fourier Space or the 
Scalar Field Power Spectrum:
\begin{equation}
    \mathcal{P}(k) = \frac{k^3|\delta\chi_k|^2}{2\pi^2}
    \label{eq:4.32}
\end{equation}

\begin{figure}[H]
    \centering
    \makebox[\textwidth][c]{%
        \includegraphics[width=0.9\textwidth]{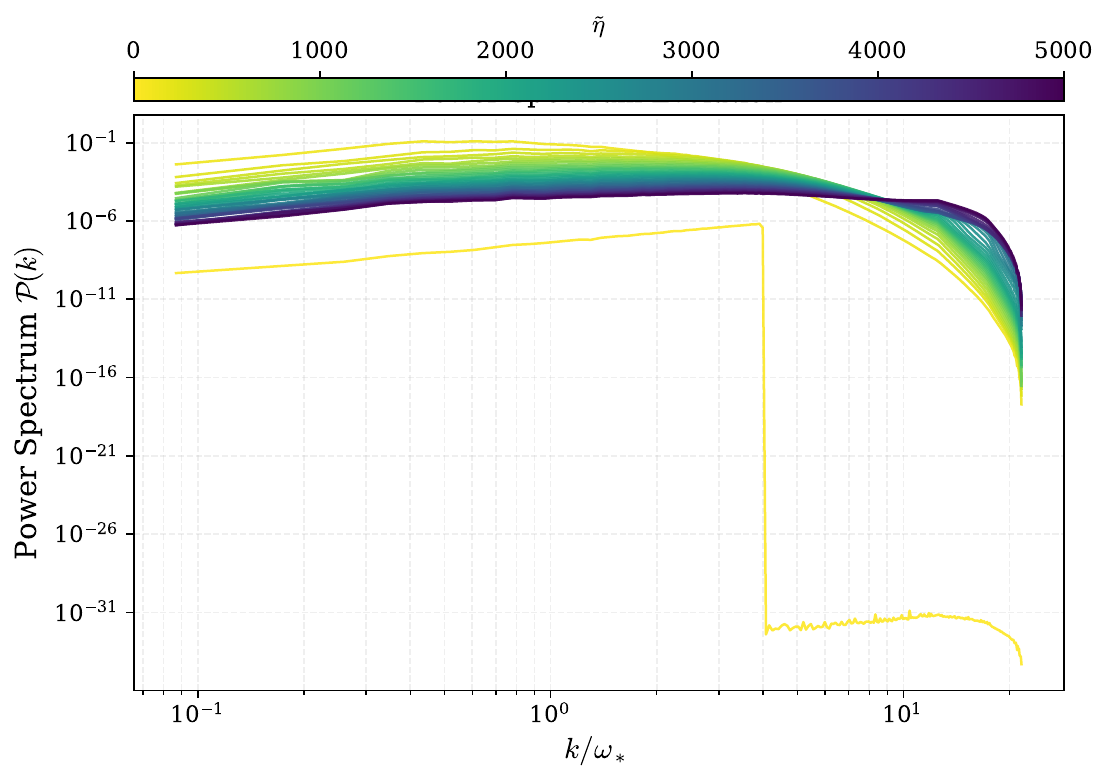}}
    \caption{Time evolution of the Scalar field power 
    spectrum, $\mathcal{P}(k) = k^3|\delta\chi_k|^2/2\pi^2$, 
    with each spectrum taken at intervals of 
    $\Delta\tilde{\eta} = 25$.}
    \label{fig:powerspectrum}
\end{figure}

The power spectrum determines how the field amplitude fluctuations are distributed across different spatial scales. From Figure~\ref{fig:powerspectrum}, we can clearly observe that for $k/\omega_* \lesssim 0.8$, there is a sharp enhancement in the perturbations, as stated in Section~\ref{sec: linear stability}. Initially, at early times, the growth of the fluctuations is confined to low momenta (infrared modes) because only modes with $k < k_{\rm tach}$ experience a negative effective mass squared and are therefore unstable, whereas the higher modes have very tiny variations in them. As time progresses, the perturbations reach a point where the non-linearities become significant and higher $k$ (UV) modes start getting populated.

\subsubsection{Energy Profile and Formation of Overdensities}
\begin{figure}[H]
    \centering
    \makebox[\textwidth][c]{\includegraphics[width=1.0\linewidth]{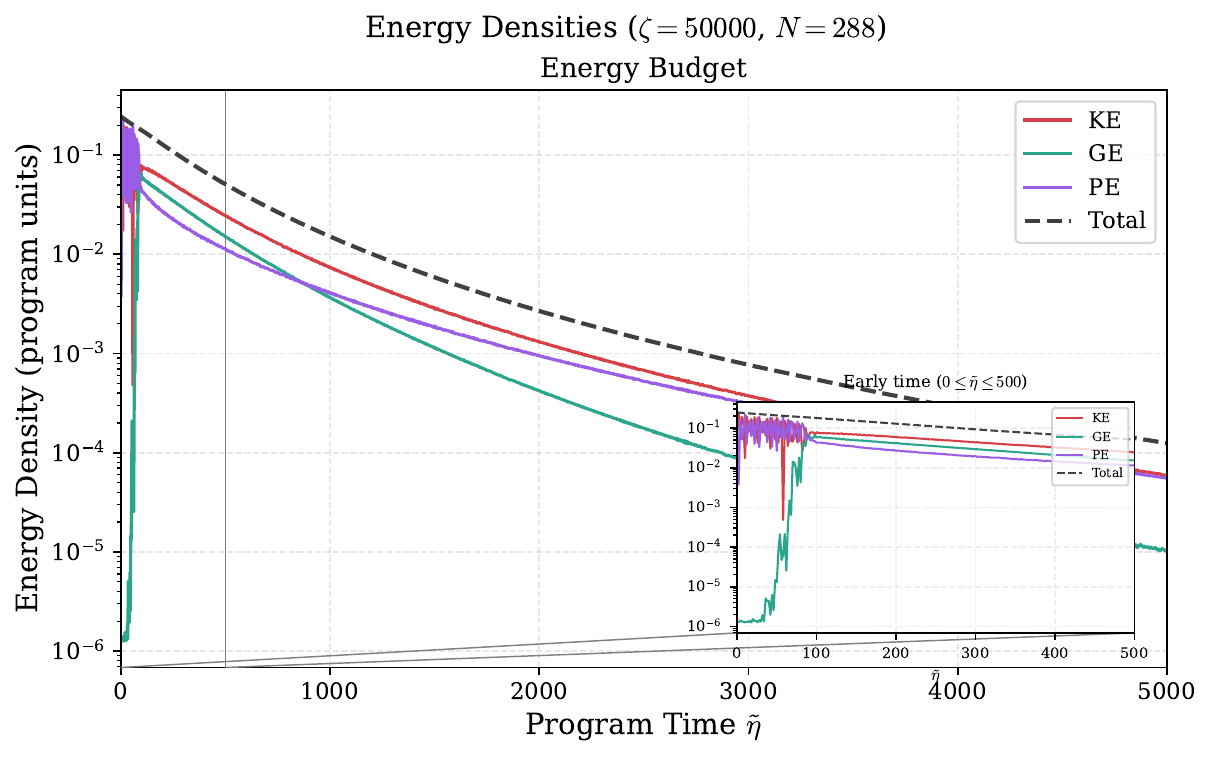}}
        \vspace{-20pt}
    \caption{Evolution of volume averaged Energies}
    \label{Energy Density Evolution}
\end{figure}

We now examine how the energy of the system evolves as the simulation progresses. We plot all three components --- kinetic, gradient, and potential. Gradient energy is the energy stored in the spatial variations of the field. From Figure~\ref{Energy Density Evolution}, it is evident that at early times of homogeneous inflaton oscillations, the total energy of the system is dominated by kinetic and potential energy. The \textit{transition} phase is characterized by a relatively short period during which the enhanced field fluctuations start to influence the oscillating inflaton condensate, albeit not sufficiently to entirely suppress the oscillations. During this stage, the gradient energy $E_G$ begins to increase significantly towards $\mathcal{O}(E_k)$ and $\mathcal{O}(E_V)$. During the \textit{backreaction} phase, the amplified field fluctuations effectively influence the background evolution, hence substantially altering its dynamics, making the inflaton's nature no longer oscillatory. This also terminates the enhancement of field fluctuations, ultimately resulting in a cessation of the increase in $E_G$.

The fact that the gradient energy remains roughly 
equipartitioned with the other energy components confirms 
that the field has permanently lost its homogeneity. The 
high gradient energy assures that the universe is filled 
with overdensities of energy at different spatial 
locations.

\begin{figure}[H]
    \centering
    \begin{subfigure}[b]{0.48\textwidth}
        \centering
        \includegraphics[width=0.9\linewidth]
        {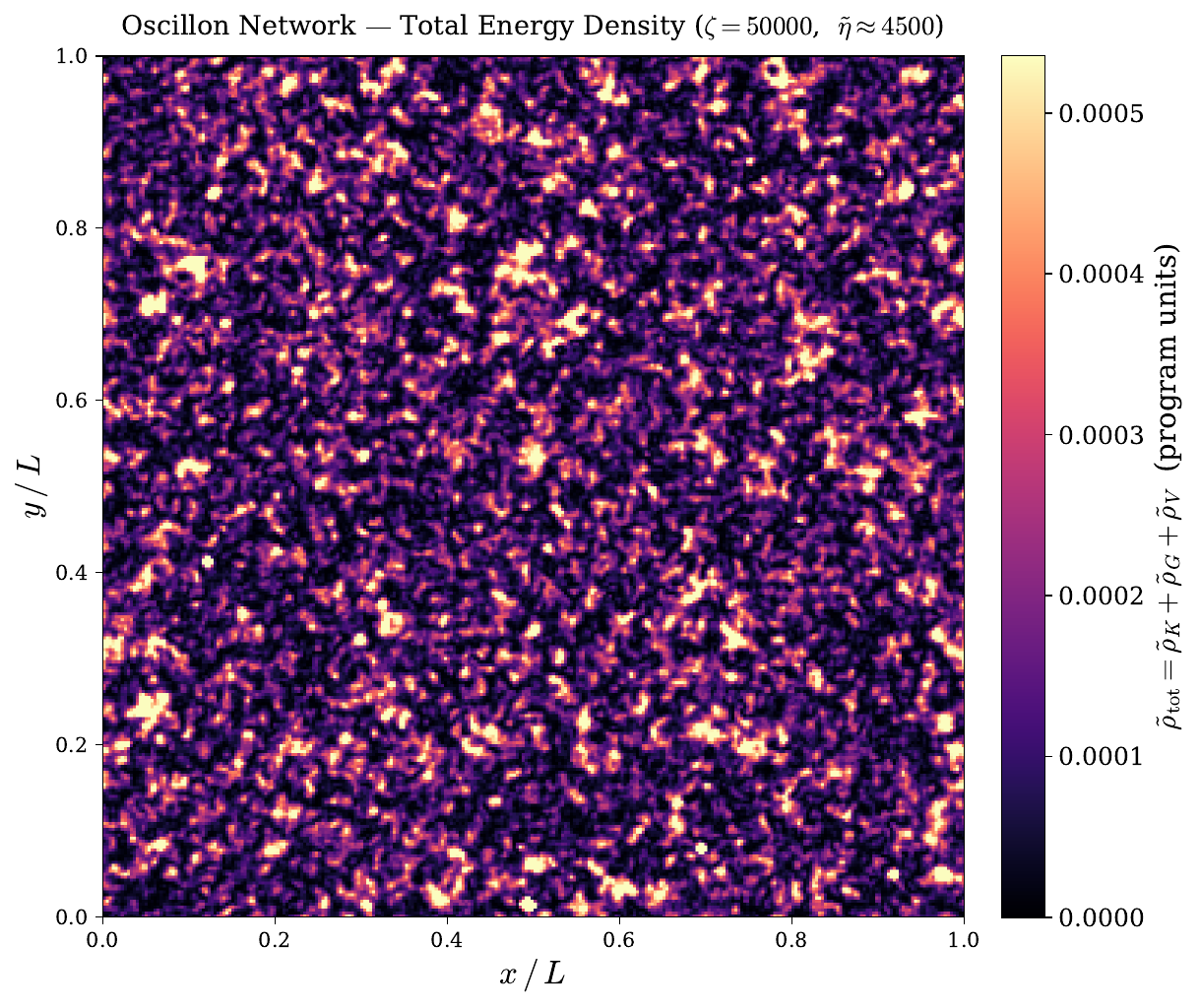}
        \caption{Absolute Total Energy Density 
        ($\tilde{\rho}_{\mathrm{tot}}$)}
        \label{fig:oscillon_absolute}
    \end{subfigure}
    \hfill
    \begin{subfigure}[b]{0.48\textwidth}
        \centering
        \includegraphics[width=0.9\linewidth]
        {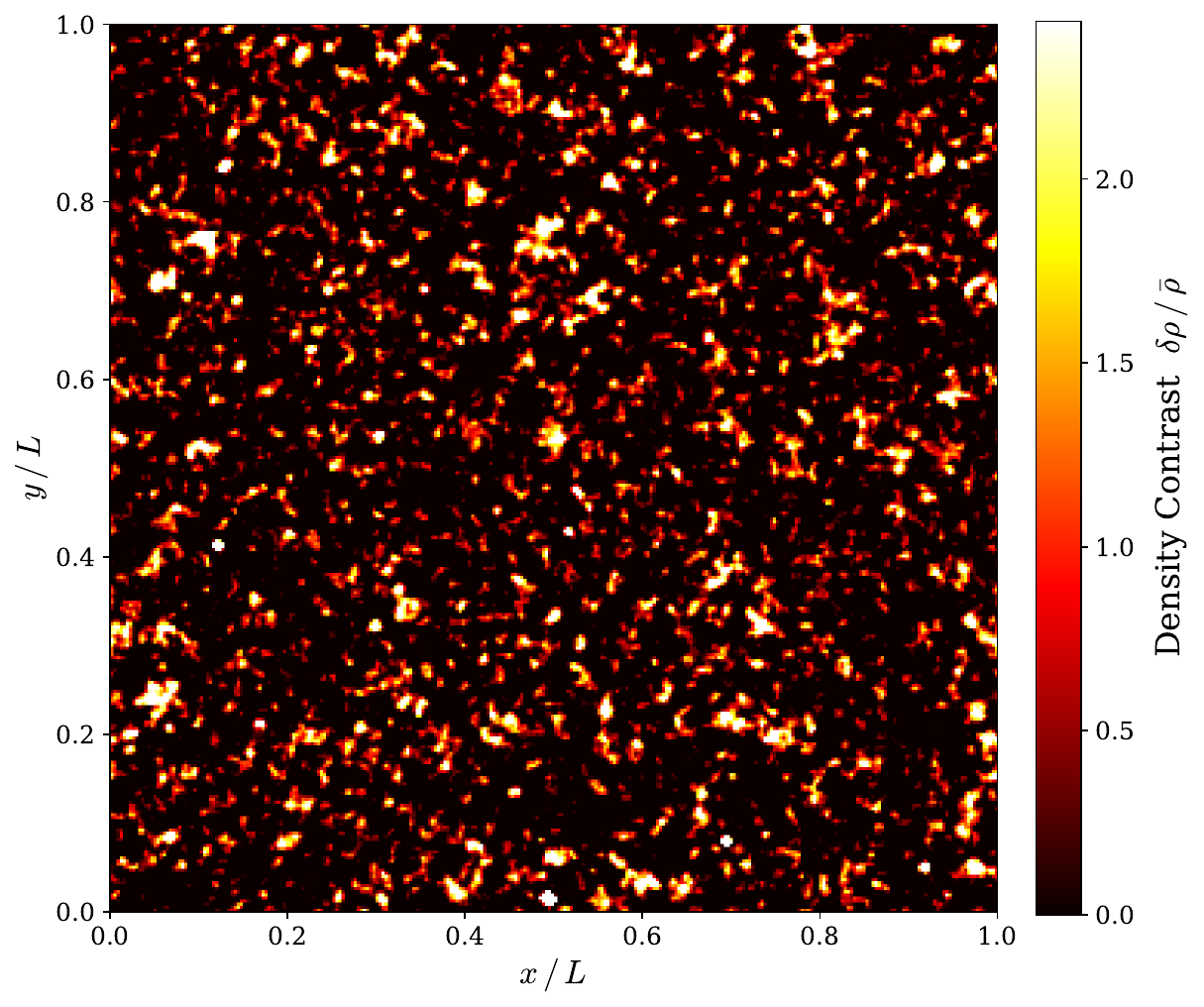}
        \caption{Total Density Contrast 
        ($\delta\rho / \bar{\rho}$)}
        \label{fig:oscillon_contrast}
    \end{subfigure}
    \caption{The spatial distribution of the Inflaton's 
    Energy Density. Panel (a) shows the absolute total 
    energy density ($\tilde{E}_{K} + \tilde{E}_{G} + 
    \tilde{E}_{V}$), representing the raw energy 
    distribution. Panel (b) shows the exact same physical 
    space plotted as a density contrast, mathematically 
    demonstrating that the oscillon cores are highly 
    localized overdensities significantly denser than the 
    background vacuum.}
    \label{fig:oscillon_comparison}
\end{figure}

The above plots distinctly illustrate field fragmentation, 
with oscillons depicted as bright glowing spots 
(hotspots) that possess a higher energy density than the 
surrounding backdrop with few isolated peaks. In 
Fig.~\ref{fig:oscillon_contrast}, the most energetic 
oscillons exhibit a density tenfold greater than that of 
the background.

\begin{figure}[H]
    \centering
    \makebox[\textwidth][c]{\includegraphics[width=0.6\textwidth]{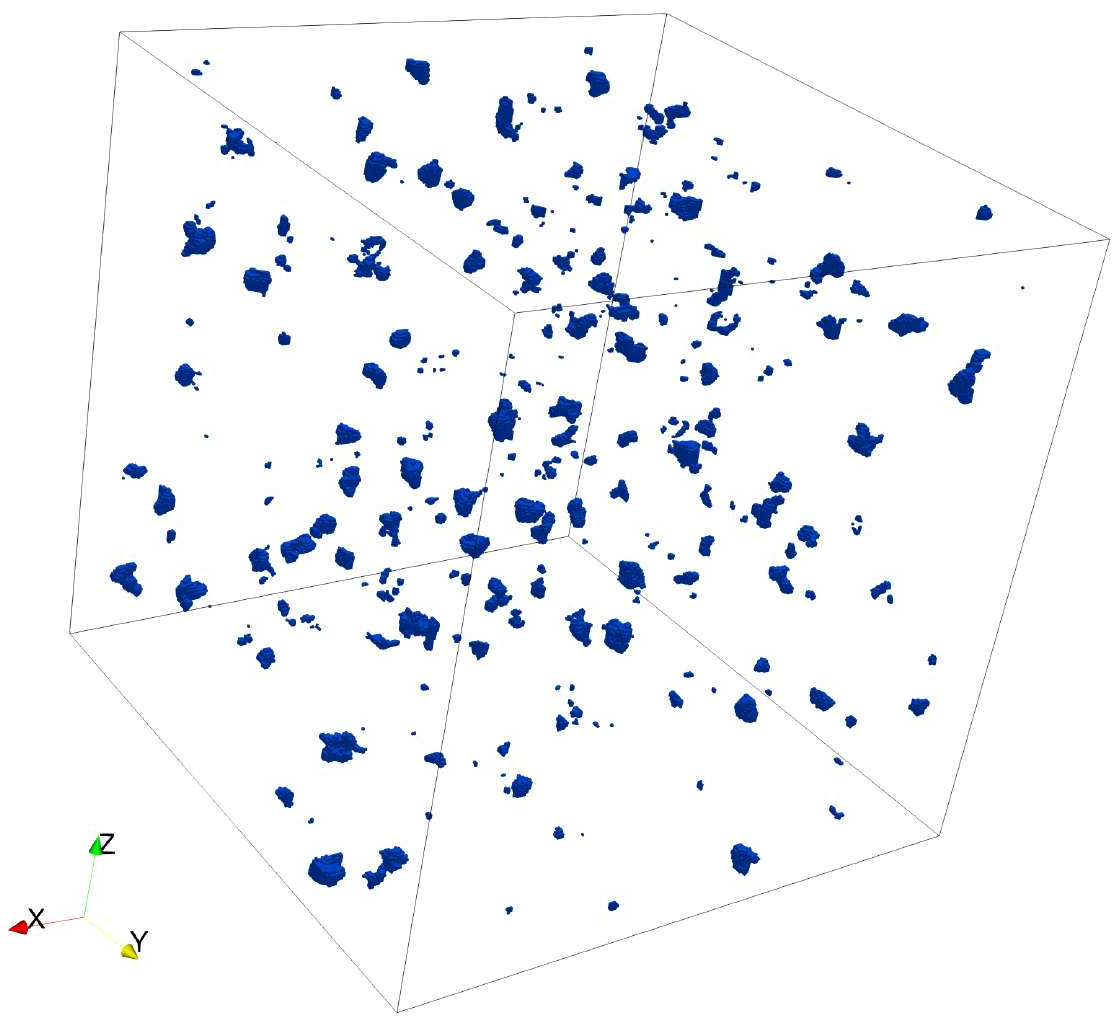}}
    \caption{3D distribution of the inflaton's energy overdensity extracted from the lattice simulation at $\tilde{\eta} = 500$
after the onset of fragmentation at $\tilde{\eta}_{\rm frag}\approx 90$, when the oscillon overdensities are most prominent. Shown are isosurfaces of the density contrast $\delta\rho/\bar{\rho} > 10$, enclosing $\sim 0.034\%$ of the simulation volume. Upon formation, these objects maintain nearly constant spatial positions for a significant number of oscillations.}
    \label{3d spatial distribution}
\end{figure} 

\subsubsection{Equation of State Parameter $\bar{\omega}$} 
We now move on to examine the behavior of the time averaged $\bar{\omega}$. In a purely homogeneous universe, assuming no particle production, the zero mode oscillates about the potential minimum as its amplitude drops adiabatically due to Hubble friction, $\Phi \propto a^{-\frac{3}{2}(1+\bar{\omega})}$ where $\bar{\omega}$ is the time averaged equation of state \cite{PhysRevD.28.1243} :
\begin{equation}
    \bar{\omega}=\frac{\langle p \rangle}{\langle \rho \rangle} \simeq\frac{n-1}{n+1}
    \label{4.33}
\end{equation}
where $n$ is determined from the potential behavior around its minimum, $\Phi^{2n}$. Hence if the potential acts as quadratic around its minimum i.e $n=1$ then we have $\bar{\omega}=0$ and the universe is matter dominated with the amplitude decreasing as $\Phi \propto a^{-3/2}$. For a potential behaving as quartic around the minimum, $n=2$, we have radiation dominated epoch with $\bar{\omega}=\frac{1}{3}$.

\begin{figure}[htbp]
    \centering
    \begin{subfigure}[t]{0.75\textwidth}
        \centering
        \includegraphics[width=\textwidth]{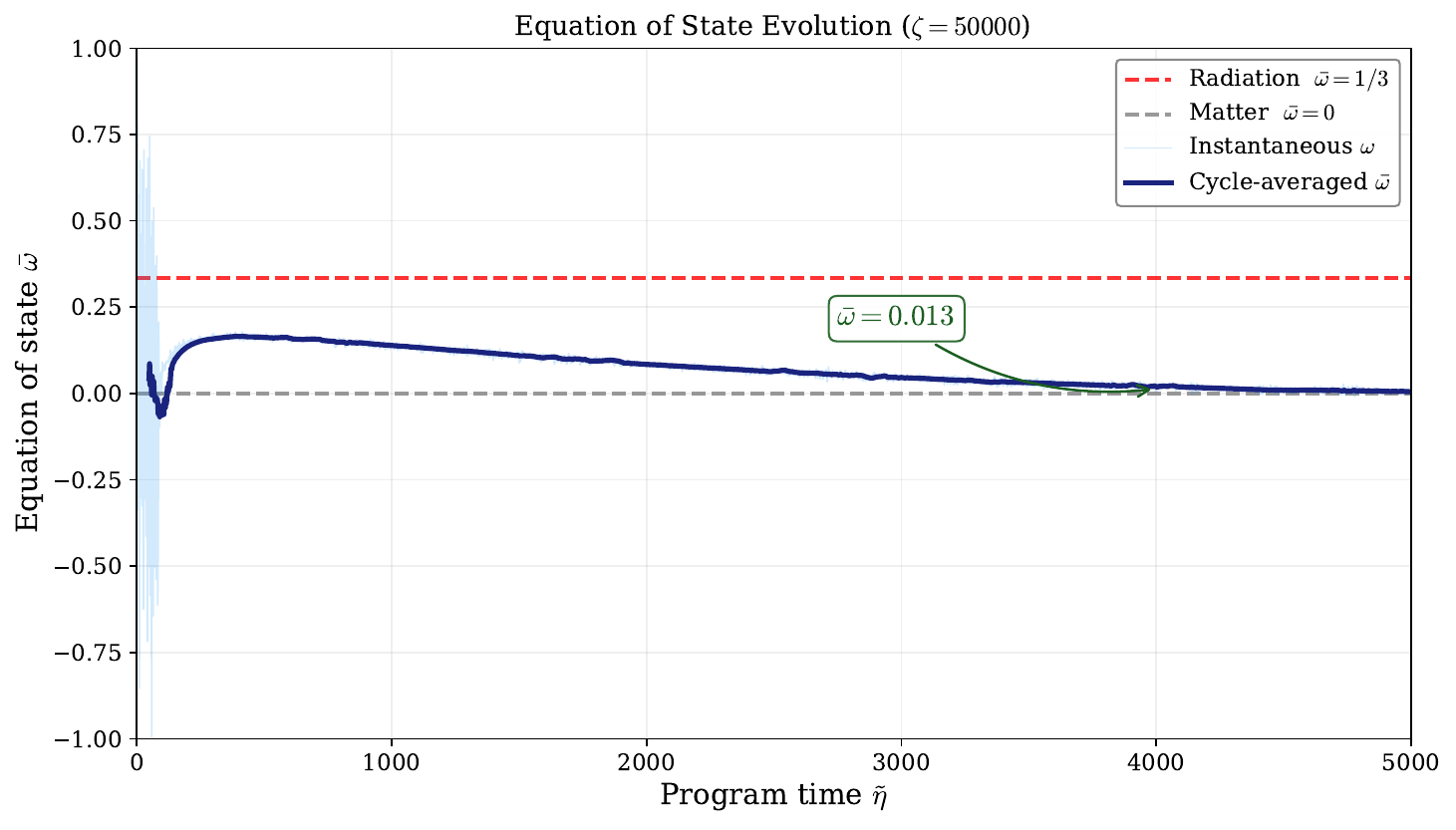}
        \caption{Time evolution of the instantaneous (light 
        blue) and cycle-averaged (dark blue) equation of state 
        parameter $\bar{\omega}$}
        \label{fig:eos_time}
    \end{subfigure}
    \hfill
    \begin{subfigure}[t]{0.75\textwidth}
        \centering
        \includegraphics[width=\textwidth]{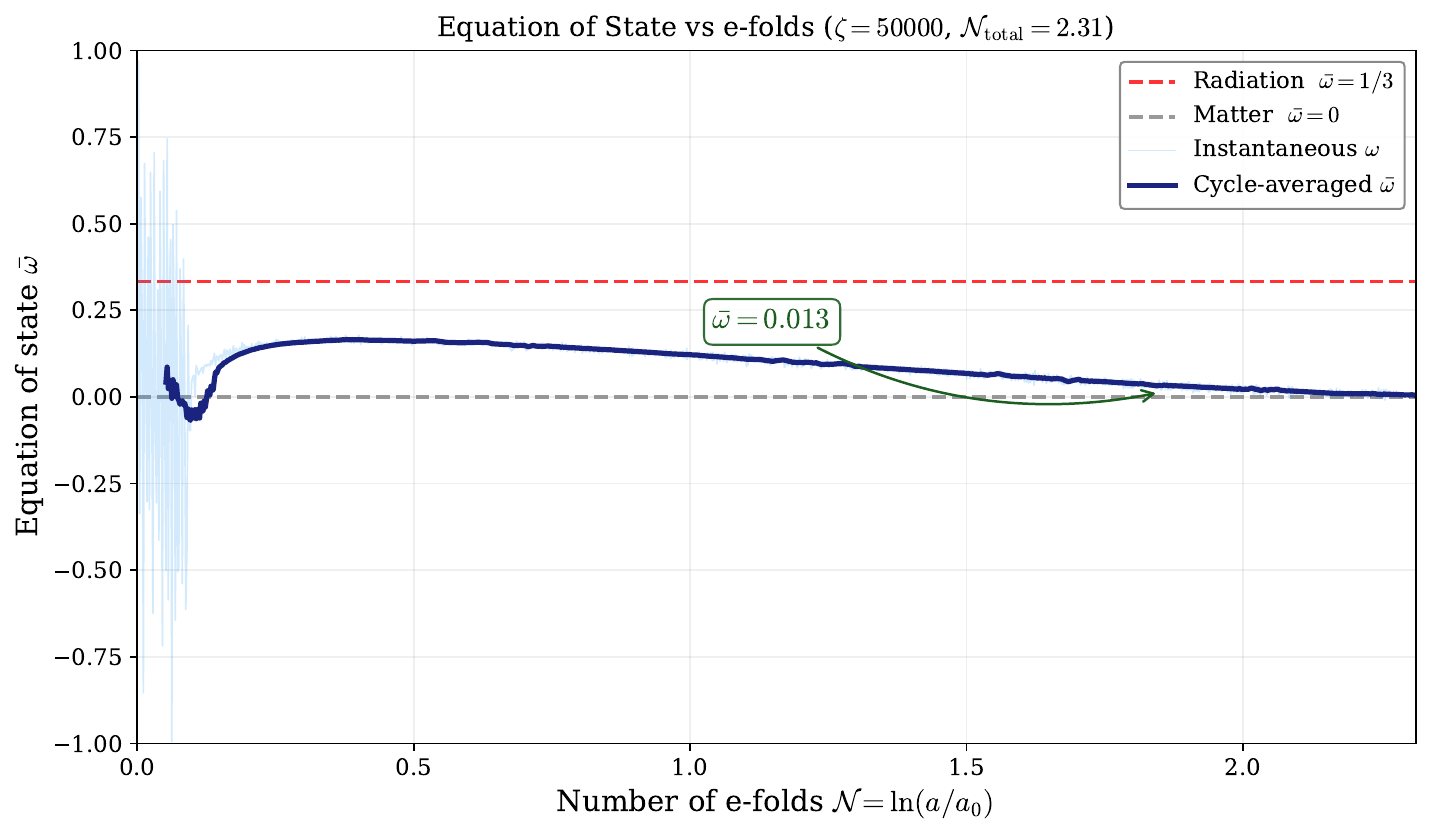}
        \caption{Equation of state parameter $\bar{\omega}$ 
        as a function of the number of e-folds 
        $\mathcal{N} = \ln(a/a_0)\approx 2.3$.}
        \label{fig:eos_efolds}
    \end{subfigure}
    \caption{Evolution of the equation of state parameter 
    $\bar{\omega}$ for $\zeta = 50000$, 
    $\alpha = 0.005$. The cycle-averaged $\bar{\omega}$ 
    stabilizes at $0.013$ at late times, demonstrating the 
    approach to matter domination ($\bar{\omega} \to 0$).}
    \label{fig:eos}
\end{figure}

The equation of state fluctuates dramatically between $-1$ and $+1$, and from \eqref{4.30}, we may deduce that these two values signify control by potential energy and kinetic energy, respectively. Initially, $\bar{\omega}$ approaches zero, indicating matter domination. It then rises to an intermediate value before decreasing monotonically toward zero over longer timescales, covering $\mathcal{N} \approx 2.13$ $e$-folds of post-inflationary evolution. The cycle-averaged equation of state stabilizes at $\bar{\omega} = 0.013$, well within the range $0 < \bar{\omega} < \frac{1}{3}$, signifying the transition between the matter-dominated epoch and the radiation-dominated epoch. This is consistent with the leading quadratic term of $U(\chi)$ near the potential minimum and in agreement with previous oscillon simulations \cite{Lozanov_2018, shafi2024formationdecayoscillonsinflation}. If the gradient energy always plays a significantly subdominant role such that $E_G \ll E_V$, then $\bar{\omega} = 0$ for $n = 1$, indicating a matter-dominated stage. 

We are presently in the preheating phase, which signifies an incomplete universe, as $\bar{\omega}$ never attains $\frac{1}{3}$. For Big Bang nucleosynthesis to occur, a radiation-dominated universe is necessary, necessitating \textbf{reheating} \cite{Kofman_1994}. To accomplish this phase, it is necessary for the inflaton to interact with other Standard Model fields, enabling energy transfer to relativistic particles (radiation) via perturbative decay channels.

\subsection{Gravitational Waves from Oscillons}
Significant scalar density perturbations, such as those linked to the oscillon number density fluctuations, can generate tensor perturbations, specifically gravitational waves, at the second order in cosmic perturbation theory. This section examines whether fragmentation and oscillon generation result in unique gravitational wave signatures. These signatures were actively investigated in the literature in various scenarios, which can be found by the reader in \cite{lozanov2026constraininginflatonpotentialgravitational,Sang_2019}. 

Three potential scenarios exist for the generation of gravitational waves in the early universe dominated by oscillons: formation, a stable oscillon phase, and decay followed by the thermalization of the universe. We will focus on the first two periods and exclude the last, as it exceeds the range of parameters of our existing model for the reasons outlined in the preceding section. In general relativity, a spherically symmetric object cannot produce gravitational waves, as it lacks a time-varying quadrupole moment ($\ddot{Q}_{ij} = 0$), which is necessary since gravitation is a spin-2 field. Therefore, a solitary spherically symmetric oscillon is incapable of emitting gravitational radiation. But it is evident from Figures~\ref{fig:oscillon_absolute} and \ref{3d spatial distribution} that the distribution of oscillons is considerably asymmetrical, and this network cumulatively generates gravitational waves. 

The substantial and dense energy blobs exert mutual gravitational attraction and exhibit relative motion, resulting in a time-varying quadrupole moment, $\ddot{Q}_{ij} \neq 0$, which serves as a key source of gravitational radiation in the early universe. These can be referred to as \textbf{primordial gravitational waves} due to their origins in the early epochs of cosmic history.

Anisotropies in the scalar field geometry can produce transverse-traceless ($TT$) linear metric perturbations $h_{ij}$, which give rise to gravitational waves, superimposed over the flat FLRW metric:
\begin{equation}
    ds^2=-dt^2+a(t)^2\left[\delta_{ij}+h_{ij}\right]dx^idx^j \hspace{0.2cm} 
    \label{4.34}
\end{equation}
Using the Transverse-Traceless gauge we have the following constraints :
\begin{equation}
    \partial_{i}h_{ij}=0 \hspace{0.2cm} \text{and} \hspace{0.3cm} h_{ii}=0
    \label{4.35}
\end{equation}
The dynamics of the metric perturbations is governed by the following wave equation :
\begin{equation}
    \ddot{h}_{ij}+3H\dot{h}_{ij}-\frac{1}{a^2}\left[\nabla^2h_{ij}+2\Pi^{TT}_{ij}\right]=0
    \label{4.36}
\end{equation}
where $\Pi^{TT}_{ij}=(\partial_i\Phi\partial_j\Phi)^{TT}$ is the Transverse-Traceless part of the anisotropic Energy-Momentum Tensor. The Gravitational Wave Energy Density Power Spectrum is given by :
\begin{equation}
    \Omega_{GW} = \frac{1}{\rho_c}
\frac{d\rho_{GW}}{d\ln(k)} \quad 
\text{with} \quad \rho_{GW} = 
\frac{1}{4}\langle\dot{h}_{ij}\dot{h}_{ij}\rangle
    \label{4.37}
\end{equation}

We make use of the \textit{Cosmo}$\mathcal{L}$\textit{attice} GW module \cite{baeza2022cosmolattice_gw} to find the gravitational wave spectrum. A thorough analytical and numerical study of gravitational waves from preheating has also been carried out in \cite{Zhou_2013}. It is crucial to remember that the gravitational waves generated during the preheating period will be redshifted if observed today; thus, we must consider the appropriate conversion factor when determining the present spectrum. This is done by adhering to the procedure outlined in \cite{Dufaux_2007}.

\begin{figure}[H]
    \centering
    \makebox[\textwidth][c]{\includegraphics[width=1.1\textwidth]{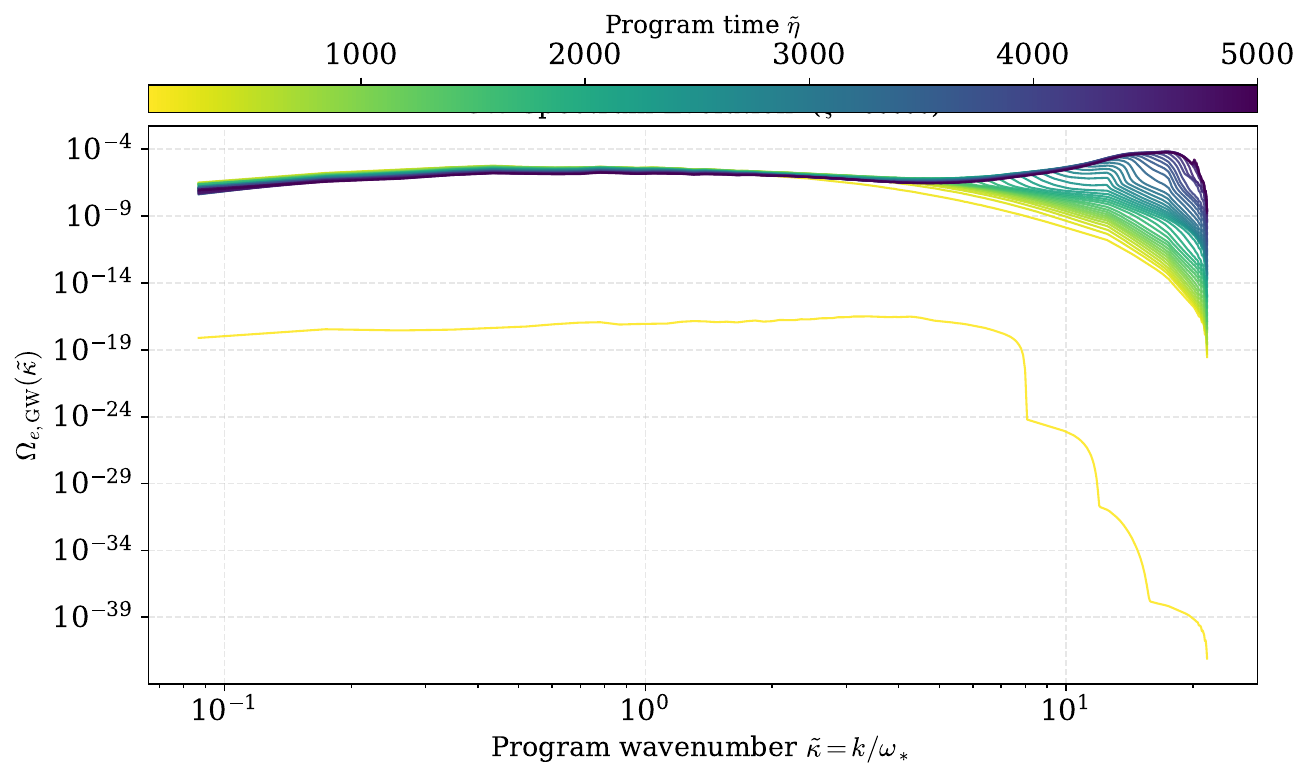}}
    \caption{Gravitational Wave Spectral Energy Density at the time of emission as a function of wavenumber}
    \label{GW Spectrum at the time of emission}
\end{figure}

Initially, we present the spectrum at the moment of emission in the early universe, followed by the computation of the current spectral energy density of gravitational waves. From Figure~\ref{GW Spectrum at the time of emission}, it is evident that at early times in the simulation, low $k$-modes generated GWs that have a larger energy density than those generated from high $k$-modes. As the system evolves in time, the energy density $\Omega_{e,GW}$ increases. Upon the onset of fragmentation, the spectrum rapidly escalates. Upon the formation of oscillons, the spectrum ceases to expand, indicating their unique nature and nearly spherical symmetry.

Ultimately, we will focus on the current abundance of gravitational wave energy density, $\Omega_{0,GW}h^2$, given by 
\begin{equation}
    h^2\big(\frac{\rho_{GW}}{\rho_c}\big)_0=\int \frac{d\nu}{\nu}h^2\Omega_{GW}(\nu)
    \label{4.38}
\end{equation}
and its spectrum :
\begin{equation}
    h^2\Omega_{0,GW}(\nu)=\big(\frac{h^2}{\rho_c}\frac{d\rho_{GW}}{d \hspace{0.3mm}ln(\nu)}\big)_0
    \label{4.39}
\end{equation}
with $\nu$ being the frequency and $\rho_c=\frac{3H_0^2}{8\pi G}$ is the critical energy density today. To convert the primordial spectrum to observables that are relevant today we need to know the evolution history of the Universe from the preheating until now, which eventually depends on the equation of state. Our simulation covering $\mathcal{N} \approx 2.13$ e-folds of post-inflationary evolution shows that the cycle-averaged equation of state stabilizes at $\bar{\omega} = 0.013$. We assume that the thermalization process occurs by perturbative reheating.

Assuming entropy conservation from the moment the universe thermalizes until today, we can suitably rescale the amplitude $\Omega_{\mathrm{GW},0}$ and frequency $f_0$ of the spectrum, resulting in expressions similar to those given in \cite{Piani_2025}: 
\vspace{0.3em}
\begin{equation}
  f_0 = \frac{k}{a_e\rho_e^{1/4}}\left[e^{-\frac{1}{4}(1-3\bar{\omega})\Delta N_{\mathrm{RD}}}\right] \hspace{0.2cm} 4\times10^{10} \mathrm{~Hz},
  \label{4.40}
\end{equation} 
and 
\begin{equation}
   \Omega_{\mathrm{GW},0} h^2 = \Omega_{\mathrm{rad},0} h^2 \left( \frac{g_{*s,0}}{g_{*s,\mathrm{RD}}} \right)^{1/3} \Omega_{\mathrm{GW},e} \times e^{ \left[ -(1 - 3\bar{\omega})\Delta N_{\mathrm{RD}} \right]}
    \label{4.41}
\end{equation}
where the subscript $e$ represents the value at the time of emission in the early universe, and $0$ represents the value today. $\Delta N_{\mathrm{RD}}$ is the number of $e$-folds between the period of oscillon domination and radiation domination. $\bar{\omega}$ is the equation of state parameter at the time of emission. $g_{*s}$ is the effective number of relativistic degrees of freedom for entropy in that particular epoch. $\Omega_{\mathrm{rad},0}h^2$ is the present-day radiation energy density, with $h$ being the uncertainty in the Hubble parameter ($h \approx 0.7$).

The parameter $\Delta N_{\mathrm{RD}}$ quantifies the number of $e$-folds of expansion between the end of the lattice simulation at $\tilde{\eta} = 5000$ and the onset of radiation domination, during which the universe is dominated by the oscillon condensate with the measured mean equation of state $\bar{\omega} = 0.013$. Since \textit{Cosmo}$\mathcal{L}$\textit{attice} evolves the scalar field on a homogeneous FLRW background and does not simulate the oscillon decay process, $\Delta N_{\mathrm{RD}}$ remains a free parameter encoding the uncertainty in the post-simulation reheating history. During this transition phase, the GW amplitude scales logarithmically per $e$-fold according to
\begin{equation}
    \frac{d\ln\Omega_{\mathrm{GW}}}{dN} = -(1 - 3\bar{\omega}),
    \label{4.42}
\end{equation}
and the peak frequency is shifted by $e^{-(1-3\bar{\omega})/4}$ per $e$-fold, both relative to the $\Delta N_{\mathrm{RD}} = 0$ case. For $\bar{\omega} = 0.013$, these factors evaluate to $e^{-0.961} \approx 0.38$ and $e^{-0.240} \approx 0.79$, respectively.

The value $\Delta N_{\mathrm{RD}} = 0$ corresponds to instantaneous reheating at the end of the simulation and represents a \emph{rigorous upper bound} on the GW signal: any additional $e$-folds during this epoch can only suppress the amplitude further. The range is truncated at $\Delta N_{\mathrm{RD}} = 3$ on the basis of the following arguments.
\begin{enumerate}

    \item \textbf{Diminishing amplitude suppression.} By $\Delta N_{\mathrm{RD}} = 3$,
    the GW amplitude has already been reduced to
    \begin{equation}
        \frac{\Omega_{\mathrm{GW}}(\Delta N_{\mathrm{RD}}=3)}
             {\Omega_{\mathrm{GW}}(\Delta N_{\mathrm{RD}}=0)}
        = e^{-3(1-3\bar{\omega})} = e^{-3\times 0.961} \approx 0.056,
        \label{4.43}
    \end{equation}
    i.e., a suppression of 94\% relative to the upper bound. For $\Delta N_{\rm RD} = 4$ and $5$, the signal falls to roughly 2.1\% and 0.8\% of the upper bound, respectively.  Beyond $\Delta N_{\mathrm{RD}} \approx 4\text{--}5$, the 
    successive curves become visually indistinguishable on a logarithmic scale, 
    and displaying them adds no physical information.

    \item \textbf{Observational irrelevance beyond $\Delta N_{\mathrm{RD}} \approx 3$.}
    The GW signal from our current model peaks at 
    $f_0 \approx 4\text{--}5\;\text{GHz}$ with a characteristic strain 
    $h_c \sim \mathcal{O}(10^{-32})$, which lies approximately ten orders of 
    magnitude below the strain sensitivity of current detectors. 
    Since the signal is undetectable across the entire range 
    $\Delta N_{\mathrm{RD}} \in \{0,1,2,3\}$, extending the range further carries 
    no observational consequence for the foreseeable future.

    \item \textbf{BBN lower bound on the reheating temperature.} Big Bang 
    nucleosynthesis (BBN) requires the universe to enter radiation domination 
    with a lower limit on the reheating temperature of $T_{\mathrm{RD}} \geq 4\;\text{MeV}$~\cite{Kawasaki_2000,
    Hannestad:2004nb}. The reheat temperature after $\Delta N_{\mathrm{RD}}$ additional 
    $e$-folds scales as
    \begin{equation}
        T_{\mathrm{RD}} \;\sim\; \rho_e^{1/4}\,
        \exp\!\left(-\frac{3(1+\bar{\omega})}{4}\,\Delta N_{\mathrm{RD}}\right),
        \label{4.44}
    \end{equation}
    where $\rho_e^{1/4} \approx 1.6\times10^{14}\;\text{GeV}$ is the energy 
    scale at the end of the simulation. Evaluating this limit for our equation of state $\bar{\omega} = 0.013$ reveals that the BBN bound of $T_{\rm RD} \geq 4$ MeV is only violated if the oscillon-dominated phase persists for $\Delta N_{\rm RD} \gtrsim 50$ additional $e$-folds. Since our analysis truncates long before this limit, the BBN constraint is entirely non-binding for the range considered in this model. It is nevertheless cited here as the standard cosmological consistency check applied in the literature~\cite{Figueroa:2017vfa,Dufaux:2007pt}.

    \item \textbf{Oscillon quasi-stability.} For the sextic-stabilized potential 
    adopted in this work ($\sigma \neq 0$), classical oscillons are quasi-stable 
    with lifetimes far exceeding a Hubble time at the relevant epoch. The quantum 
    decay rate is suppressed by the bounce action $S_E \sim 1/\lambda \sim 10^7$, 
    rendering perturbative decay negligible. Consequently, $\Delta N_{\mathrm{RD}}$ is 
    genuinely uncertain and could in principle be large. However, since the 
    observational imprint of additional $e$-folds saturates by $\Delta N_{\mathrm{RD}} 
    \approx 3\text{--}4$, extending the range serves no purpose.
\end{enumerate}

We now check the observability of these gravitational waves generated during the oscillon-dominated epoch by computing the spectral energy density today against the Power-Law Integrated Sensitivity Curves (PLICs) for current and future GW observatories. PLICs represent detector sensitivity curves for stochastic gravitational-wave backgrounds, accounting for the enhanced sensitivity derived from both frequency and temporal integration. These PLICs incorporate a boost in sensitivity derived from the broadband characteristics of the signal, achieved through frequency integration. To construct the PLICs for various detectors from detector noise power spectral density, transfer functions, and overlap reduction functions, we follow the procedure from \cite{PhysRevD.88.124032}.

\begin{figure}[H]
    \centering
    \makebox[\textwidth][c]{%
        \includegraphics[width=1.1\textwidth]
        {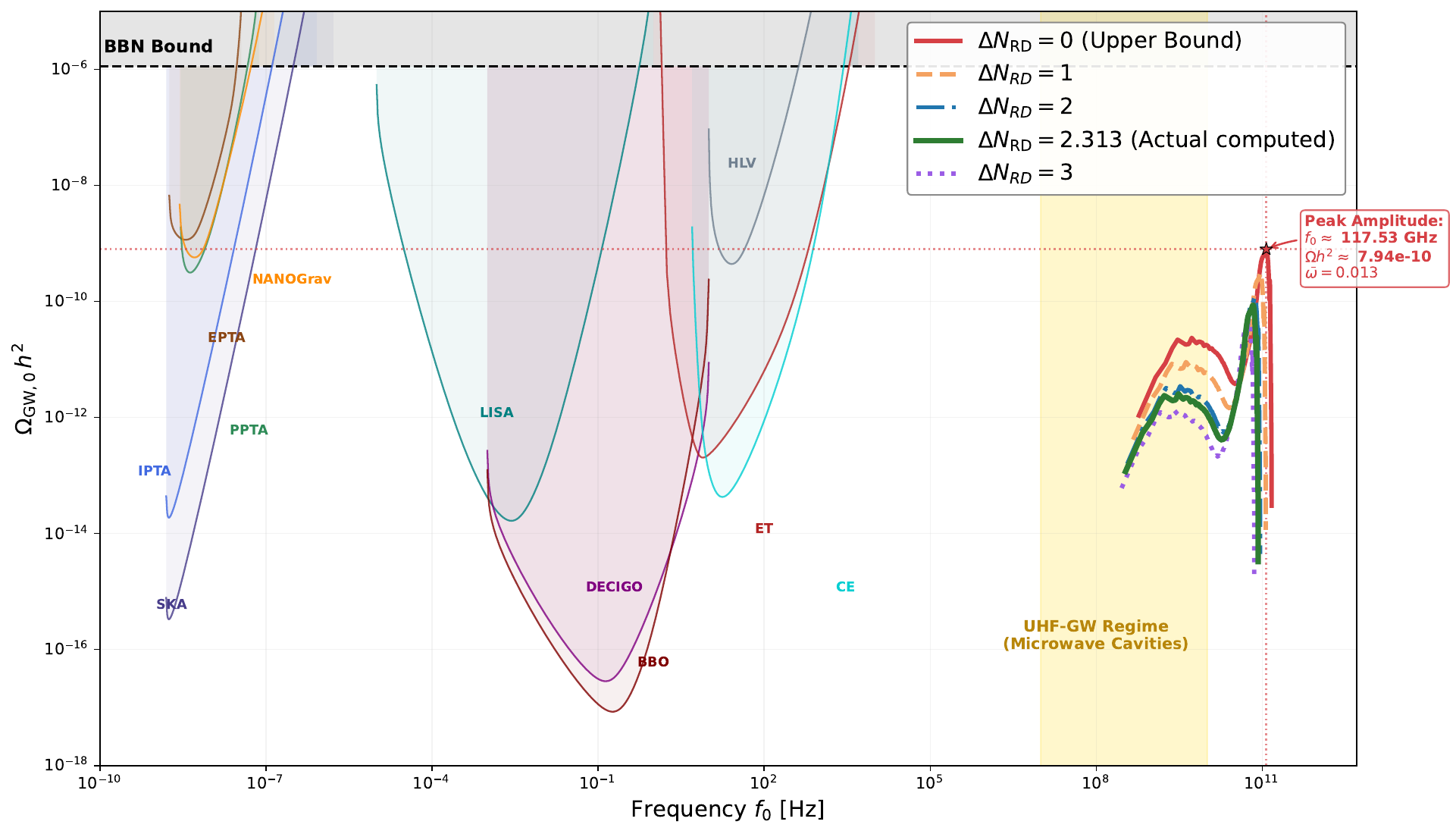}}
    \caption{Spectral Energy Density of Gravitational 
    Waves and Frequency, generated from Oscillons as 
    observed today for $\Delta N_{RD} = 0, 1, 2, 3$ 
    plotted against the PLICs for various detectors. 
    The gray region above highlights the BBN bound.}
    \label{fig:gwspectrum}
\end{figure}
\newpage
The above plot display the PLICs for various 
detectors as shaded backgrounds, alongside the spectrum 
from oscillon-generated gravitational waves represented 
by solid curves for different values of 
$\Delta N_{\mathrm{RD}}$. The four curves 
$\Delta N_{\mathrm{RD}} \in \{0,\,1,\,2,\,3\}$ 
therefore span the complete physically and 
observationally meaningful uncertainty band on the GW 
prediction, with the additional curve 
$\Delta N_{\mathrm{RD}} = 2.13$ representing the actual 
value computed from our simulation covering 
$\mathcal{N} \approx 2.13$ e-folds of post-inflationary 
evolution with a measured equation of state 
$\bar{\omega} = 0.013$. The curve labeled 
$\Delta N_{\mathrm{RD}} = 0$ constitutes the upper bound 
on the signal, while successive curves illustrate the 
progressive suppression due to a prolonged matter-like 
phase prior to thermalization. 
Figure~\ref{fig:gwspectrum} indicates that, despite the 
enhanced spectral amplitude of the primordial spectrum 
observed today, the peak occurs at $\approx 117.53$ GHz, 
which is beyond the detection range of present 
gravitational wave experiments. Despite the current 
detectors operating at a frequency range many orders of 
magnitude lower, numerous strategies are being devised 
to explore ultra-high-frequency signals from the early 
universe inside the MHz--GHz spectrum 
\cite{Aggarwal2021, Aggarwal2025, 
amaral2026globaldetectornetworksearch}.

The GW spectrum exhibits two distinct features, reminiscent 
of the double-peak structure observed in other oscillon-forming 
models \cite{2021JHEP...03..021H, Zhou_2013}. The dominant peak at 
$f_0 \approx 117.53$ GHz arises from the broad tachyonic 
instability during inflaton fragmentation, with the peak 
frequency set by the characteristic oscillon scale $\sim m^{-1}$. 
The secondary peak at lower frequencies is consistent with 
the internal frequency content of the oscillons \cite{2021JHEP...03..021H}, which 
generates a power deficiency between the two peaks. While the precise origin of 
this secondary peak in our modified gravity model would 
require a dedicated analysis beyond the scope of this work, 
its presence within the UHF-GW regime accessible to resonant 
microwave cavity experiments \cite{Berlin_2022} 
provides an additional observational signature of oscillon 
formation.\\ 
The sensitivity curves for BBO, CE, ET, DECIGO, LISA, 
PTAs, SKA, and NANOGrav are obtained from 
\cite{Schmitz_2021}, while those for the resonant 
microwave cavities are from \cite{PhysRevD.108.124009}. 
The PLICs for ASTROD-GW and $\mu$Ares were obtained from 
their respective noise spectra given in 
\cite{doi:10.1142/S0218271813410046} and 
\cite{Sesana2021}. Similar construction of all these 
curves was also done in \cite{Kuralkar:2025hoz}.

One of the proposed Gravitational Wave Experiments \cite{hong2026highsensitivitymethodologiesdetect} that makes use of Pulsars and next generation of Radio Telescopes like FAST and SKA, have computed the observable characteristic strain $h_c$ for Gravitational Waves from Preheating era to be of the order of $10^{-32}- 10^{-33}$ and the frequency of the order of Gigahertz (upto $\approx 100 GHz$).
This experiment would act as a good cross verification to test the signals of Stochastic Gravitational Wave Background from early universe during (P)reheating epoch and also from Oscillons. 

\begin{figure}[H]
    \centering
    \makebox[\textwidth][c]{\includegraphics[width=1.2\textwidth]{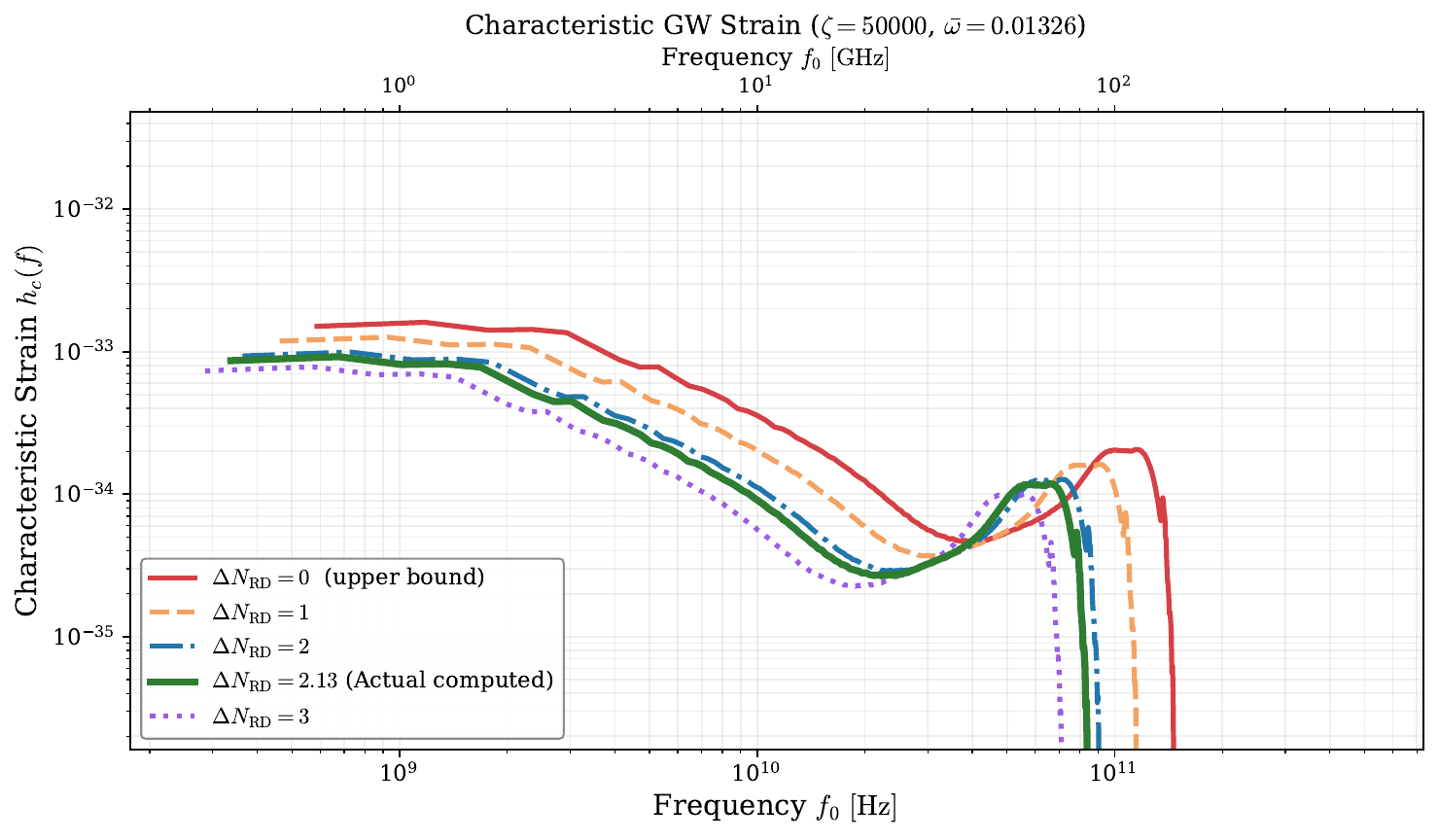}}
    \caption{Characteristic Strain of Gravitational Waves vs Frequency}
    \label{GW strain}
\end{figure}

We can in-fact verify this by computing the characteristic strain for our model and plotting that against the frequency in figure \ref{GW strain}.
The characteristic strain $h_c(f)$ is derived by connecting the 
gravitational wave energy density to the metric perturbation through 
the relation
\begin{equation}
    \langle h^2 \rangle = \int_0^{\infty} h_c^2(f)\,d\ln f\,,
    \label{4.45}
\end{equation}
one can combine both the equations in ~\eqref{4.37}
to obtain
\begin{equation}
    h_c(f) = \sqrt{\frac{3H_0^2}{2\pi^2}\,
    \frac{\Omega_{\rm GW,0}(f)\,h^2}{f^2}}
    \;\approx\; 
    1.263\times10^{-18}
    \left(\frac{1\;{\rm Hz}}{f}\right)
    \sqrt{\Omega_{\rm GW,0}(f)\,h^2}\
    \label{4.46}
\end{equation}
where $H_0 = h\times 100\;{\rm km\,s^{-1}\,Mpc^{-1}}$ is the 
present Hubble rate and the uncertainty in it is $h \simeq 0.7$

\section{Summary and Discussion}
\label{Sec: summary}
In this paper, we have established a detailed framework to study oscillons, which are spatially localized, oscillatory in time, pseudo-stable excitations (energy configurations) of the inflaton scalar field formed during the preheating stage of the universe, just after inflation. We considered a modified gravity model in the Palatini formalism of general relativity, featuring a non-minimal coupling of the scalar field with gravity, with a coupling function of the form $f(R,\Phi)$. We consider an oscillatory, polynomial (sextic) potential with an attractive non-linear term, which satisfies the conditions required for the formation of oscillons. Subsequently, we introduced the strategy of stability analysis and Floquet theory to observe which momentum modes actually grow owing to tachyonic instability and eventually give rise to oscillons. We implemented the numerical technique of $3+1$ classical lattice simulations using \textit{Cosmo}$\mathcal{L}$\textit{attice}. It was observed that when the fluctuations become comparable to the mean field values, the inflaton fragments and loses its oscillatory nature, with the increase in gradient energy being one of the reasons, along with the tachyonic instability. For our particular choice of model parameters, the power spectrum reveals that low-momentum modes are excited at early times, and a sharp enhancement peak is observed for $k \lesssim 0.8m$. The cycle-averaged value of the equation of state, $\bar{\omega}$, after rendering close to $0$ at initial times, stabilizes to a constant value throughout the remaining time of the simulation.

Additionally, we have demonstrated that these structures can generate a substantial gravitational wave signal, offering a potential observational avenue for this inflationary model. We conclude that the amplitude is commensurable with the sensitivity of current GW experiments, but the frequency is $\sim \mathcal{O}(\text{GHz})$ (ultra-high-frequency regime), far away from those of present-day detectors. However, the characteristic strain and frequency lie well within the range of proposed future gravitational wave detectors.

This work constitutes a preliminary effort toward understanding oscillon dynamics in preheating scenarios; nonetheless, there are some elements that our analysis did not adequately address, along with approximations utilized, which require mention:
\begin{enumerate}
    \item \textbf{Oscillon Lifetime:} Knowing the lifespan of oscillons is crucial for accurately assessing the inflationary observables to be compared with empirical data. Furthermore, oscillons may produce gravitational waves upon collapse, with extended lifetimes correlating to increased probabilities of detecting gravitational waves within the frequency ranges of existing and future gravitational wave observatories. Additionally, if sufficiently stable, oscillons may ultimately cluster over extended time periods \cite{PhysRevD.100.063507}.

    \item \textbf{Gravitational Effects and Interactions:} Given that oscillons are well-localized entities with significant overdensities, it is reasonable to question whether these quasi-spherical formations could plausibly collapse into primordial black holes \cite{9965-5h5h}, a scenario not considered in our simulations, which entirely neglect metric perturbations. Our methodology models the late-time behavior of the system by considering the oscillons as isolated entities, hence disregarding any possible interactions that could induce deformations or subtle alterations in the energy profile.

    \item \textbf{Towards a Reheating Mechanism:} A minimal amount of radiation is necessary post-inflation to reheat the universe. Our model does not inherently result in radiation domination from the depletion of the scalar field's energy density to levels below that of radiation. To facilitate reheating, we can offer a framework guided by \cite{Kofman_1994} and followed by \cite{Kuralkar:2025hoz}. Sub-dominant vacuum fluctuations grow significantly post-inflation, resulting in particle production. The particles created at the completion of inflation alter the Lagrangian by incorporating interaction terms between the scalar field and the particles. These interaction factors are additionally accountable for the disintegration of the scalar field into those particles. This decay reduces the energy density of the field, facilitating the necessary radiation domination for effective reheating.
\end{enumerate}

\section*{\Large{Appendix A}} \label{Appendix A}
\setcounter{equation}{0}
\renewcommand{\theequation}{A.\arabic{equation}}
\textbf{\LARGE{Matrix Monodromy Method}}

\vspace{0.5em}
This method \cite{Hauser_2001}  is standard technique implemented while solving Hill's Differential Equation in matrix form as given in equation \eqref{3.17}. It is the $2\times2$ linear system with coefficient matrix being periodic, $U(t+T)=U(t)$. The matrix $U(t)$ has two notable properties :
\begin{itemize}
    \item $tr(U)=0$
    \item $U$ is real
\end{itemize} 
A general solution of the matrix system can be proposed by the following ansatz :
\begin{equation}
    \delta(t)=M(t)\delta(0)
    \label{A.1}
\end{equation}
where $M(t)$ is the fundamental matrix solution often called the \textbf{propagator matrix} satisfying the following :
\begin{equation}
    \dot{M}(t)=U(t)M(t) \hspace{0.4cm} \text{with} \hspace{0.4cm} M(0)=1
    \label{A.2}
\end{equation}
The propagator carries all the information about the evolution of any initial state.
By Liouvilles Theorem, we have the determinant of $M(t)$ conserved as $tr(U)=0$ from which we obtain :
\begin{equation}
    \frac{d|M(t)|}{dt}=tr(U)\cdot |M(t)|=0
\end{equation}
giving the constraint equation : 
\begin{equation}
    det\hspace{0.2mm} M(t)=det \hspace{0.3mm} M(0)=det \hspace{0.3mm}I=1 \hspace{0.4cm} \forall \hspace{0.5mm}t
    \label{A.3}
\end{equation}
This implies the two eigen values are always reciprocals of each other.

The \textbf{Monodromy Matrix} is the propagator evaluated at exactly one full period :
\begin{equation}
    \textbf{M}\equiv \textbf{M}(T)=
\begin{pmatrix}
M_{11} & M_{12} \\
M_{21} & M_{22}
\end{pmatrix}
\label{A.5}
\end{equation}
with $det\hspace{0.3mm}M=M_{11}M_{22}-M_{12}M_{21}=1$ 

Due to periodicity, after $n$ full periods we get :
\begin{equation}
    \textbf{M}(nT) = \textbf{M}(T)^n=\textbf{M}^n
    \label{A.6}
\end{equation}
Using the initial conditions \eqref{3.20} we can construct the full Monodromy matrix as :
\begin{equation}
M =
\begin{pmatrix}
\delta \chi_k^{(1)}(T) & \delta \chi_k^{(2)}(T) \\
\delta \pi_k^{(1)}(T)  & \delta \pi_k^{(2)}(T)
\end{pmatrix}
\label{A.7}
\end{equation}
Simultaneously integrating both the solutions as a $4 \times4$ system :
\begin{equation}
\frac{d}{dt}
\begin{pmatrix}
\delta \chi^{(1)} \\
\delta \pi^{(1)} \\
\delta \chi^{(2)} \\
\delta \pi^{(2)}
\end{pmatrix}
=
\begin{pmatrix}
0 & 1 & 0 & 0 \\
-\omega_k^2 & 0 & 0 & 0 \\
0 & 0 & 0 & 1 \\
0 & 0 & -\omega_k^2 & 0
\end{pmatrix}
\begin{pmatrix}
\delta \chi^{(1)} \\
\delta \pi^{(1)} \\
\delta \chi^{(2)} \\
\delta \pi^{(2)}
\end{pmatrix}.
\label{A.8}
\end{equation}
\textbf{Floquet Multipliers} : Eigen values of $\textbf{M}$. Floquet theorem states that if $\delta(0)$ is an eigen vector of $\textbf{M}$ with eigenvalue $\lambda_m$, then : 
\begin{equation}
\boldsymbol{\delta}(nT) = \mathbf{M}^n \boldsymbol{\delta}(0) = \lambda^n \boldsymbol{\delta}(0)
\label{A.9}
\end{equation}
The eigenvalues $\lambda_{m\pm}$ are called \textit{Floquet multipliers} with charactersitic equation as :
\begin{equation}
\lambda_m^2 - (M_{11} + M_{22})\,\lambda_m + 1 = 0
\label{A.10}
\end{equation}
\text{The solutions are:}
\begin{equation}
\lambda_{m\pm}
=
\frac{M_{11} + M_{22}}{2}
\pm
\sqrt{
\left( \frac{M_{11} + M_{22}}{2} \right)^2 - 1
}
\label{A.11}
\end{equation}
The Floquet exponent can be extracted as :
\begin{equation}
    \mu_k=\frac{1}{T}ln(\lambda_m)
    \label{A.12}
\end{equation}
Taking the real part, we obtain
\begin{equation}
\mathcal{R}(\mu_k^{\pm}) 
= \frac{1}{T} \ln \left|
\frac{M_{11} + M_{22}}{2}
\pm
\sqrt{
\left( \frac{M_{11} + M_{22}}{2} \right)^2 - 1
}
\right|.
\label{A.13}
\end{equation}

In terms of the solution components, this becomes
\begin{equation}
\mathcal{R}(\mu_k^{\pm}) 
= \frac{1}{T} \ln \left|
\frac{\delta \chi_k^{(1)}(T) + \delta \pi_k^{(2)}(T)}{2}
\pm
\sqrt{
\frac{\left[\delta \chi_k^{(1)}(T) - \delta \pi_k^{(2)}(T)\right]^2}{4}
+ \delta \chi_k^{(2)}(T)\,\delta \pi_k^{(1)}(T)
}
\right|.
\label{A.14}
\end{equation}
\bibliographystyle{JHEP}
\bibliography{main}
\end{document}